\documentclass[12pt,a4paper]{article}
\input epsf
\usepackage{amsmath,amsfonts,epsfig,color,latexsym}
\usepackage{amssymb}
\usepackage[english, USenglish]{babel}
\usepackage{epsfig}
\usepackage{a4wide}
\usepackage{rotating}
\usepackage{slashed}

\newcommand{\pa}{\partial}
\newcommand{\be}{\begin{equation}}
\newcommand{\ee}{\end{equation}}
\newcommand{\bea}{\begin{eqnarray}}
\newcommand{\eea}{\end{eqnarray}}

\renewcommand{\a}{\alpha}

\newcommand{\da}{{\dot\alpha}}
\newcommand{\ad}{{\dot a}}
\newcommand{\bd}{{\dot b}}

\renewcommand{\b}{\beta}

\renewcommand{\l}{\lambda}

\newcommand{\q}{\theta}
\newcommand{\bq}{\bar\theta}
\newcommand{\bu}{\bar u}

\newcommand{\ep}{\epsilon}
\newcommand{\m}{\mu}
\newcommand{\n}{\nu}

\newcommand{\cN}{{\cal N}}

\newcommand{\bt}[1]{{\bar t}}

\newcommand{\eisen}[1]{\mathcal{P}_{#1}}

\newcommand{\coup}[2]{\mathcal{A}_{#1#2}}
\newcommand{\couph}[2]{\mathcal{A}_{#1#2}^{\text{het}}}

\newcommand{\Gsum}{\sum_{(P^L,P^R)\in\Gamma^{(6,22)}}}
\newcommand{\GsumT}{\sum_{(P^L,P^R)\in\Gamma^{(2,2)}}}
\newcommand{\qq}{q^{\frac{1}{2}(P^L)^2}\bar{q}^{\frac{1}{2}(P^R)^2}}

\author{\\[-0.3cm]\Large I.~Antoniadis\footnote{\tt ignatios.antoniadis@cern.ch}
\footnote{On leave from CPHT (UMR CNRS 7644) Ecole Polytechnique, F-91128 Palaiseau}~\,and S.~Hohenegger\footnote{{\tt stefanh@itp.phys.ethz.ch}}
}
\title{\begin{flushright}{\vspace{-0.8cm}\small CERN-PH-TH/2009-162}\end{flushright}
\vspace{0.8cm}
\bf{$\cN=4$ Topological Amplitudes and Black Hole Entropy}}
\dateUSenglish
\date{}
\graphicspath{{figures/}}
\begin{document}
\begin{titlepage}
\maketitle
\begin{center}
\renewcommand{\thefootnote}{\fnsymbol{footnote}}\vspace{-0.5cm}
\footnotemark[1]Department of Physics, CERN - Theory Division, CH-1211 Geneva 23, Switzerland\\[0.5cm]
\vspace{-0.3cm}
\footnotemark[3]Institut f\"ur Theoretische Physik, ETH
  Z\"urich, 
CH-8093 Z\"urich, Switzerland\\[0.5cm]
\vspace{2cm}
\end{center}
\begin{abstract}
We study the effects of $\cN=4$ topological string amplitudes on the entropy of black holes. We analyse the leading contribution associated to six-derivative terms and find one particular operator which can correct the entropy of $\cN=4$ black holes. This operator is BPS-like and appears in the effective action of 
type~II string theory on $K3\times T^2$ or equivalently its heterotic dual on $T^6$. In both descriptions the leading contribution arises at one-loop, which we calculate explicitly on the heterotic side. We then consider whether this term has any consequences for the entropy of (large) $\cN=4$ black holes and find that it makes indeed a contribution at subleading order. Repeating the computation for small black holes with vanishing horizon area at the classical level, we prove that this coupling lifts certain flat directions in the entropy function thereby being responsible for the attractor equations of some moduli fields.
\end{abstract}
\thispagestyle{empty}
\end{titlepage}

\tableofcontents
\renewcommand{\theequation}{\arabic{section}.\arabic{equation}}
\section{Introduction}
BPS-type interactions have over the years attracted a lot of attention in four-dimensional extended supergravity. These are couplings which can be written as integrals over a subspace of the full superspace thereby generalising the notion of chirality and F-terms in $\cN=1$ supersymmetric field theories. Within the effective string theory action, such terms are believed to be always captured by topological amplitudes; the best studied case is indeed the series of the $g$-loop couplings $F_gW^{2g}$ in type II string theory compactified on a Calabi-Yau manifold \cite{Antoniadis:1993ze,Bershadsky:1993cx}. Here $W$ is the chiral $\cN=2$ supergravity multiplet and the moduli-dependent coefficient function $F_g$ was shown to be identical to the genus $g$ partition function of the $\cN=2$ topological string, associated to the twisted Calabi-Yau $\sigma$-model. 

Among many interesting properties of $F_g$'s, it was realised that they play an important role for the physics of supersymmetric black holes. In \cite{Lopes Cardoso:1998wt,LopesCardoso:1999cv,Lopes Cardoso:1999ur,LopesCardoso:1999xn} (see also \cite{Mohaupt:2000mj}) higher derivative corrections to the entropy have been derived from these effective action terms, following a method first proposed in \cite{Wald:1993nt}. These results -- at least for large values of the charges of the black hole -- are in agreement with state-counting arguments in a microscopic description of the black hole as a particular configuration of branes (see e.g.~\cite{Maldacena:1997de,Vafa:1997gr}). Similar results have more recently been found even for particular $\cN=4$ supersymmetric small black holes, as for example in \cite{Dabholkar:2004yr}. There, a D0-D4-brane setup has been studied in type~II string theory compactified on $K3\times T^2$. It was shown that the only non-vanishing coupling from the series $F_{g=1}$ for the case of $\cN=4$ supersymmetry, which is a four-derivative operator, yields the full entropy of the black hole and agrees to all orders in the large D0-D4 brane charge expansion with the expected result from microstate counting.

In \cite{Ooguri:2004zv} an even more direct link between $F_g$ and $\cN=2$ black holes was established by conjecturing a relation of the form $Z_{\text{BH}}=|Z_{\text{top}}|^2$. Here $Z_{\text{BH}}$ is the ``thermodynamic'' partition function of the black hole in a particular mixed ensemble and $Z_{\text{top}}$ is essentially the exponential of the weighted sum over all $F_g$'s. This conjecture is understood to hold perturbatively, since a non-perturbative definition of either side of the equality is generically unclear. A somewhat deeper understanding of this relation (particularly for the square on the right hand side) was reached in \cite{Beasley:2006us}. Moreover, the conjecture has been tested for small supersymmetric black holes in \cite{Dabholkar:2005dt,Dabholkar:2005by}.

The results mentioned so far raise the question whether generalisations of $F_g$ to theories with $\cN=4$ supersymmetry have a similar impact on the physics of four-dimensional $\cN=4$ supersymmetric black holes. Such generalisations have first been found in \cite{Antoniadis:2006mr} in type~II string theory compactified on $K3\times T^2$ (see also \cite{Berkovits:1994vy}). Explicitly, two series of higher derivative BPS couplings have been identified both of which are computed by certain correlation functions of the $\cN=4$ topological string: $\mathcal{F}_g^{(1)}\bar{K}^2 K^{2g}$ and $\mathcal{F}_{g-1}^{(3)} K^{2g}$, where $K$ is a superdescendant of the $\cN=4$ supergravity multiplet. Particularly the latter coupling was extensively studied in \cite{Antoniadis:2007cw} (see also \cite{Antoniadis:2007ta,Antoniadis:2009nv}) for values $g\geq2$. In this work we will mostly be concerned with the expression for $g=1$, which corresponds to a six-derivative operator. Using string dualities, we will see that this coupling starts receiving contributions at one-loop in heterotic string theory compactified on $T^6$, which we can therefore study fairly explicitly. 

We will then carry on to determine the effect of $\mathcal{F}_{g-1}^{(3)}$ with $g=1$ on the entropy of certain $\cN=4$ supersymmetric black holes. The method we will apply is the classical entropy function formalism developed in \cite{Sen:2005wa,Sen:2005iz} (for a review see e.g.~\cite{Sen:2007qy}). This is a suitable approach to the problem as it does not necessitate the knowledge of the complete solution of the black hole in the presence of the higher derivative terms, but nevertheless it allows to extract information about the near horizon geometry and most importantly the corrected entropy of the black hole. We should also mention that our approach is `classical' in the sense that non-local terms arising from integrating out massless degrees of freedom are not included. We should also point out that we have made a general analysis of dimension six operators and we found one more candidate, BPS-like on-shell involving three Riemann tensors, which however does not change the entropy of $\cN=4$ black holes.

This paper is organised as follows. In Section~\ref{Sect:EffectAction} we discuss a manifestly supersymmetric formulation of the couplings $\mathcal{F}_{g-1}^{(3)}$ for $g=1$ in $\cN=4$ harmonic superspace. After introducing our conventions we will show how to write these terms in an off-shell supersymmetric manner. We also prove that this coupling contains at the component level a term of the form $R_{(+)}^2F_{(-)}^2$ with $R_{(+)}$ the self-dual piece of the Riemann tensor and $F_{(-)}$ the anti-self-dual field strength tensor of a vector  multiplet gauge field. In Section~\ref{StringOneLoop} we explicitly extract the leading string theory contribution to this component interaction from a one-loop amplitude in heterotic string theory compactified on $T^6$. We compute the corresponding amplitude explicitly in a particular region of the moduli space, including the integral over the modular parameter of the world-sheet torus. We also show that a similar contribution for the gauge fields replaced by graviphotons vanishes identically. This Section is accompanied by three appendices containing additional material as well as calculations which we omitted from the main body of the paper for pedagogical reasons. In Section~\ref{Sect:EntropyBlackLarge} we use the precise form of the one-loop expression to determine its contribution to the entropy of a particular large $\cN=4$ supersymmetric black hole. We find a contribution of the order $-2$ in the charges. Repeating a similar analysis for certain small black holes in Section~\ref{Sect:EntropyBlackSmall} reveals that the entropy stemming from $R_{(+)}^2F_{(-)}^2$ is still suppressed with respect to the contributions of $R^2$ couplings. However, our six-derivative term can be shown to be responsible for the lifting of certain flat moduli directions in the entropy function, thereby providing attractor values for some scalar fields. Finally, Section~\ref{Sect:Conclusions} contains our conclusions.
\section{$\cN=4$ Supersymmetric Effective Action}\label{Sect:EffectAction}
\setcounter{equation}{0}
In this Section we discuss a particular class of higher derivative couplings of the $\cN=(4,4)$ type~II effective action, which have first been discovered in \cite{Antoniadis:2006mr,Antoniadis:2007cw}. Due to the high amount of supersymmetry, a covariant formulation of these couplings is not possible in standard superspace; for this reason we will work in harmonic superspace, for which we will first review our conventions.
\subsection{$\cN=4$ Supergravity and Harmonic Superspace Description}\label{Sect:Superfields}
In this work we will deal with black holes in $\cN=4$ Poincar\'e supergravity (SUGRA) \cite{Cremmer:1977tt,Bergshoeff:1980is,de Roo:1984gd} being the low energy limit of type~II string theory compactified on $K3\times T^2$ or its dual heterotic string theory on $T^6$. The field content of this theory is the $\cN=4$ supergravity multiplet coupled to $22$ $\cN=4$ vector multiplets. The scalar fields together form the moduli space
\begin{align}
\mathbb{M}=\frac{SU(1,1)}{U(1)}\times \frac{SO(6,22)}{SO(6)\times SO(22)}\,.
\end{align}
The $SO(6,22)$ symmetry is linearised by introducing six additional vector multiplets that
act as compensators for various (gauge-)symmetries of the theory. For example, as explained in \cite{Antoniadis:2007cw}, the 36 scalar fields of these multiplets are eliminated by imposing the D-term constraints (20 constraints) and gauge fixing Weyl invariance (one constraint) as well as the local $SO(6)$ symmetry (15 constraints). Concerning the gauge fields there are two possibilities: Either the gauge fields of the compensating multiplets are expressed as functions of the graviphotons which sit inside the supergravity multiplet ('superstring basis') or the relation is inverted and the graviphotons are 
identified with the gauge fields of the compensating multiplets; in this case, the vector bosons of the supergravity multiplet are
expressed as functions of all vector multiplet gauge fields ('supergravity basis'). Throughout this paper we will consistently work in the superstring basis which is most suitable for our purpose of calculating higher derivative couplings in string theory.

A description of this theory in standard $\cN=4$ superspace
\begin{align}
\mathbb{R}^{(4|4)}=\{x^\mu,\theta^i_\alpha,\bar{\theta}_i^{\dot{\alpha}}\}\,,\label{standardSS}
\end{align}
where $i=1,\ldots,4$ an index of $SU(4)$ (the automorphism group of $\cN=4$ supersymmetry in four dimensions),
turns out to be difficult. In fact it is only possible on-shell since the necessary superfields cannot be introduced in a consistent off-shell fashion. We will therefore choose a different description in four-dimensional {\it harmonic superspace} \cite{Galperin:1984av,Galperin:1984bu,Howe:1995md,Hartwell:1994rp}. The latter is an enhancement of (\ref{standardSS}) of the following type
\begin{align}
\mathbb{HR}^{(4+4|4)}=\mathbb{R}^{(4|4)}\times \frac{SU(4)}{S(U(2)\times U(2))}=\{x^\mu,\theta^i_\alpha,\bar{\theta}_i^{\dot{\alpha}},u^{+a}_i, \, u^{-\ad}_i\}\,.
\end{align} 
The coordinates which parameterise the additional coset space $\{u^{+a}_i, u^{-\ad}_i\}$ transform as fundamentals under $SU(4)$ and carry indices $a, \ad = 1,2$ of $SU(2)\times SU(2)$ as well as $U(1)$ charges $\pm 1$. Together with their complex conjugates $\bu^i_{+a} = \overline{(u^{+a}_i)}, \, \bu^i_{-\ad} = \overline{(u^{-\ad}_i)}$  they satisfy the unitarity conditions
\begin{align}
&u^{+a}_i\, \bu^i_{+b} = \delta^a_b\,, && u^{-\ad}_i\, \bu^i_{-\bd} = \delta^\ad_\bd \,, && u^{+a}_i\, \bu^i_{-\bd} = u^{-\ad}_i\, \bu^i_{+b} = 0\,, &&u^{+a}_i\, \bu^j_{+a} + u^{-\ad}_i\, \bu^j_{-\ad} = \delta^j_i\,,\label{12'}
\end{align}
and the unit determinant condition
\begin{align}\label{12}
    \ep^{ijkl} u^{+a}_i u^{+b}_j u^{-\ad}_k u^{-\bd}_l = \ep^{ab}\ep^{\ad\bd}\,.
\end{align}
It is furthermore convenient to introduce vector-like combinations of $SU(4)$ harmonics (i.e. harmonics on $SO(6)/SO(4)\times SO(2)$) of the type $u^M_{ij} = -u^M_{ji}$, with $M= (++,--, a\ad)$ (and their conjugates $\bu_{M}^{ij} = \overline{u^{M}_{ij}}$)
\begin{align}
& u^{++}_{ij} = u^{+a}_i \ep_{ab} u^{+b}_j\,, && u^{--}_{ij} = u^{-\ad}_i \ep_{\ad\bd} u^{-\bd}_j \,, && u^{a\ad}_{ij} = u^{+a}_{[i} u^{-\ad}_{j]}\ ,  \label{00}
\end{align}
where $[ij]$ denotes weighted antisymmetrisation. 

The introduction of harmonic variables allows us to define ``1/2-BPS short'' or Grassmann
(G-)analytic superfields.\footnote{For more details on their construction see e.g. \cite{Antoniadis:2007cw}.} They depend only on half of the Grassmann variables which can
be chosen to be $\q^{+a}_\a = \q^i_\a\, u_i^{+a}$ and $\bq^\da_{-\ad} =
\bu^i_{-\ad}\,\bq^\da_i $. One such superfield is the linearised  on-shell vector multiplet (we only display the bosonic degrees of freedom)
\begin{align}
Y_A^{++}(x^\mu,\q^+,\bq_-,u) =& \phi^{ij}_A u^{++}_{ij} + \q^{+a}\sigma^{\m\n}\q^{+b}\ep_{ab}\, F_{(+),A,\m\n} + \bq_{-\ad}\bar\sigma^{\m\n}\bq_{-\bd}\ep^{\ad\bd}\, F_{(-),A,\m\n} + \ldots\ , \label{01}
\end{align}
where the dots stand for additional derivative terms. Moreover, $\sigma^{\mu\nu}$ and $\bar\sigma^{\m\n}$ are the 4-dimensional (anti-) chiral Lorentz generators, $\phi^{ij}= \frac{1}{2}\ep^{ijkl} \bar\phi_{kl}$ are six real scalars and $F_{(\pm)\m\n}$ is the (anti-)self-dual part of the gauge field strength. Finally, we have also included the $SO(22)$ index $A$.

Another example of a G-analytic superfield is the linearised on-shell Weyl multiplet. It is obtained from the off-shell chiral Weyl superfield \cite{Bergshoeff:1980is} (we only display the bosonic degrees of freedom)
\begin{align}
\mathcal{W}=&\ \Phi+\theta^i_\alpha\theta^j_\beta\left(\sigma^{\alpha\beta}_{\mu\nu}T^{\mu\nu}_{(+)[ij]}+\epsilon^{\alpha\beta}S_{(ij)}\right)+\frac{1}{12}\epsilon_{ijkl}(\theta^i\sigma^{\mu\nu}\theta^j)(\theta^k\sigma^{\rho\tau}\theta^l)R_{\mu\nu\rho\tau}+\ldots\,.\label{N4SUGRA}
\end{align}
Here $\Phi$ is a physical scalar (``graviscalar"), $T$ is a sixplet of graviphoton field strengths, $S_{(ij)}$ is an auxiliary field and $R_{\mu\nu\rho\tau}$ is the Riemann tensor. From $\mathcal{W}$ we can compute the following superdescendant
\begin{equation}\label{16}
K^{++}_{\m\n} = (\sigma_{\m\n})_{\a\b}D_{-\ad}^\a D_{-\bd}^\b\ep^{\ad\bd}\ \mathcal{W}=(\sigma_{\m\n})_{\a\b}\ep^{\ad\bd}\bu_{-\ad}^i \bu_{-\bd}^jD^\a_i D^\b_j\mathcal{W}\,,
\end{equation}
which similarly to the vector superfield (\ref{01}) only depends on half of the $\q$ variables:
\begin{equation}\label{17}
K^{++}_{\m\n}(\q^+,\bq_-,u) = T_{(+)\m\n}^{ij}  u^{++}_{ij}  + \q^{+a}\sigma^{\l\rho}\q^{+b}\ep_{ab}\,  R_{(+)\m\n\l\rho} + \bq_{-\ad}\bar\sigma^\l \sigma_{\m\n} \sigma^\rho \bq_{-\bd}\ep^{\ad\bd}\,  \pa_\l \pa_\rho \Phi + \ldots\ .
\end{equation}
Repeating the same steps, but this time starting with the antichiral superfield $\bar{\mathcal{W}}(\bq)$ we obtain the other half of the on-shell Weyl multiplet. It is again described
by an ultrashort superfield of the same type,
\begin{equation}\label{18}
\bar K^{++}_{\m\n}(\q^+,\bq_-,u) = T_{(-)\m\n}^{ij}   u^{++}_{ij} + \bq_{-\ad} \bar\sigma^{\l\rho}\bq_{-\bd}\ep^{\ad\bd}\, R_{(-)\m\n\l\rho} + \q^{+a}\sigma^\l\bar\sigma_{\m\n} \bar\sigma^\rho \q^{+b}\ep_{ab} \,   \pa_\l \pa_\rho \bar\Phi + \ldots\,.
\end{equation}
Note that in the $\cN=4$ G-analytic superspace there exists a special conjugation $\
\widetilde{}\ $ combining complex conjugation with a reflection on the harmonic coset,
such that G-analyticity is preserved. In this sense $Y^{++} = \widetilde{Y^{++}}$ and
$\bar K^{++} = \widetilde{K^{++}}$, which implies, in particular, the reality condition
on the six scalars in $Y$.

We have now all ingredients to formulate higher order effective action couplings.
\subsection{Higher Derivative Effective Action Term}
Using the harmonic superspace approach outlined in the previous section we can construct the following higher order effective action term 
\begin{align}
S_g=\int d^4x \int du \int d^4 \q^+\int d^4\bq_- (D_-\cdot D_-)^2\left[\left(K_{\mu\nu}^{++} K^{++\,\mu\nu}\right)^g \mathcal{F}_g(\mathcal{W},Y_A^{++},u)\right]\,,\label{OffshellR2F2}
\end{align}
where we have used the shorthand notation $(D_-\cdot D_-)_{\mu\nu}=(\sigma_{\mu\nu})_{\alpha\beta}\epsilon^{\dot{a}\dot{b}}D^\alpha_{-\dot{a}}D^\beta_{-\dot{b}}$. On shell (i.e. if $S_{(ij)}=0$ in (\ref{N4SUGRA})) the only possibility to distribute the spinor derivatives is to hit two different $\mathcal{W}$ superfields inside $\mathcal{F}_g(\mathcal{W},Y_A^{++},u)$, which makes (\ref{OffshellR2F2}) equivalent to
\begin{align}
S_g=\int d^4x \int du \int d^4 \q^+\int d^4\bq_- \left(K_{\mu\nu}^{++} K^{++\,\mu\nu}\right)^{g+1} \mathcal{F}^{(3)}_g(\mathcal{W},Y_A^{++},u)\,,\label{OnshellR2F2}
\end{align}
where we have defined
\begin{align}
\mathcal{F}^{(3)}_g(\mathcal{W},Y_A^{++},u)=\frac{\partial^2 \mathcal{F}_g(\mathcal{W},Y_A^{++},u)}{\partial \mathcal{W}^2}\,.\label{N4TopDefOld}
\end{align}
The effective coupling (\ref{OnshellR2F2}) has first been considered in \cite{Antoniadis:2006mr}, given as a $(g+1)$-loop component amplitude of type~II string theory compactified on $K3\times T^2$, involving two Riemann tensors, two graviscalars with two derivatives each, and $2g-2$ graviphotons. In fact it was shown there that this amplitude is identical to a particular correlation function in the $\cN=4$ topological string, which was further studied in \cite{Antoniadis:2007cw}. Although the works \cite{Antoniadis:2006mr,Antoniadis:2007cw} focused on $g>0$ such that the above component amplitude is well defined, the case $g=0$ is also a valid contribution as can be seen from (\ref{OnshellR2F2}). In fact, in a component notation it contains among others the following term
\begin{align}
S_{g=0}^2&=\int d^4x\int du\int d^4\theta^+\int d^4\bar{\theta}_-(K^{++}_{\mu\nu} K^{++,\mu\nu}) Y_A^{++}Y_B^{++}\left(\frac{\partial^4\mathcal{F}_0(\mathcal{W},Y_A^{++},u)}{\partial Y_A^{++}\partial Y_B^{++}\partial \mathcal{W}^2}\right)\nonumber\\
&\simeq\int d^4x R_{(+),\mu\nu\rho\tau}R_{(+)}^{\mu\nu\rho\tau} F_{(-),A,\sigma\lambda} F_{(-),B}^{\sigma\lambda} \int du\,{\mathcal{A}^{AB}}_{\big|\theta=0}+\ldots\,.\label{A2compExp}
\end{align}
In the second line we have explicitly performed the Grassmann integration. To be precise, the $\theta^+$-integral has picked $R_{(+)}$ in both of the $K^{++}_{\mu\nu}$ superfields while the $\bar{\theta}_-$-integral has extracted $F_{(-)}$ from the vector multiplets. The dots denote further terms containing fermionic fields which will be of no interest when we apply (\ref{A2compExp}) to the computation of the black hole entropy. Moreover, in order to save writing we have introduced the shorthand notation
\begin{align}
\mathcal{A}^{AB}(\mathcal{W},Y_A^{++},u)\equiv \left(\frac{\partial^4\mathcal{F}_0(\mathcal{W},Y_A^{++},u)}{\partial Y_A^{++}\partial Y_B^{++}\partial \mathcal{W}^2}\right)\,.\label{ShorthandCoupling}
\end{align}
As we can see, this component term is of six derivative order. In the remainder of this work we will study the effective action coupling (\ref{A2compExp}) in more detail in order to understand whether it yields any non-trivial corrections to the entropy of black holes.
\section{String Theory One-Loop Amplitude}\label{StringOneLoop}
\setcounter{equation}{0}
As a first step we would like to study (\ref{A2compExp}) in string theory. As already mentioned, $\mathcal{F}^{(3)}_g$ in (\ref{N4TopDefOld}) has been computed as a $(g+1)$-loop string amplitude in type~II theory compactified on $K3\times T^2$, via the correlator $\langle R_{(+)}^2(\partial\partial\bar{\Phi})^2T_{(+)}^{2g-2}\rangle_{g+1}$ for $g>0$. This, however, does not smoothly connect to the coupling (\ref{A2compExp}), which is why $\coup{A}{B}$ must be computed separately. Na\"ive extrapolation suggests, however, that the latter starts receiving corrections at the one-loop level on the type~II side. Following now the steps of reasoning as in \cite{Antoniadis:2006mr} the same conclusion should in fact also be true for its heterotic dual, which we will now compute explicitly.
\subsection{One-Loop Gauge-Field Amplitude in Heterotic String Theory}\label{Sect:OneLoopHeterotic}
We consider (\ref{A2compExp}) as a one-loop amplitude in heterotic string theory compactified on $T^6$, which we will subsequently write as $T^4\times T^2$ (for similar computations see e.g. \cite{Antoniadis:1995zn,Antoniadis:1997zt,Lerche:1999ju}). The moduli of this theory are arranged in a $\Gamma^{(6,22)}$ Narain lattice, for which we will consider the simplest case, namely that none of the Wilson lines in the right moving (bosonic string) part are switched on. 
\subsubsection{Vertex Operators and Contractions}
The one-loop amplitude we need to compute contains two self-dual Riemann tensors and two anti-self-dual gauge field strengths. We choose a complex basis for the space-time (Euclidean) coordinates $(Z^1,\bar{Z}^1,Z^2,\bar{Z}^2)$ as well as their fermionic partners $(\chi^1,\bar{\chi}^1,\chi^2,\bar{\chi}^2)$. In this basis we pick the following kinematic structure for the vertices\\
\begin{center}
\begin{tabular}{|c|c|c|c|}\hline
\textbf{field} & \textbf{helicity} & \textbf{vertex} & \textbf{WS position}\\\hline
graviton & $R_{1212}$ & \parbox{6.9cm}{\vspace{0.1cm}$V_{(R)}(p_1)=\left(\partial Z^2-ip_1\chi^1\chi^2\right)\bar{\partial} Z^2e^{ip_1Z^1}$\vspace{0.1cm}} & $x_1$\\ \hline
graviton & $R_{\bar{1}\bar{2}\bar{1}\bar{2}}$ & \parbox{6.9cm}{\vspace{0.1cm}$V_{(R)}(\bar{p}_2)=\left(\partial \bar{Z}^1-i\bar{p}_2\bar{\chi}^2\bar{\chi}^1\right)\bar{\partial} \bar{Z}^1e^{i\bar{p}_2\bar{Z}^2}$\vspace{0.1cm}} & $x_2$\\ \hline
gauge field & $F_{A,\bar{1}2}$ & \parbox{6.6cm}{\vspace{0.1cm}$V^{(F)}_A(\bar{p}_1)=\left(\partial Z^2-i\bar{p}_1\bar{\chi}^1\chi^2\right)\bar{J}_A e^{i\bar{p}_1\bar{Z}^1}$\vspace{0.1cm}} & $x_3$\\\hline
gauge field & $F_{B,1\bar{2}}$ & \parbox{6.6cm}{\vspace{0.1cm}$V^{(F)}_B(\bar{p}_2)=\left(\partial Z^1-i\bar{p}_2\bar{\chi}^2\chi^1\right)\bar{J}_Be^{i\bar{p}_2\bar{Z}^2}$\vspace{0.1cm}} & $x_4$\\\hline
\end{tabular}
\end{center}
${}$\\[10pt]
where the last column denotes the position on the world-sheet. The correlator which we now have to compute is
\begin{align}
\couph{A}{B}=\langle V_{(R)}(p_1)V_{(R)}(\bar{p}_2) V^{(F)}_A(\bar{p}_1)V^{(F)}_B(\bar{p}_2)\rangle\,.\label{1LoopCorrelator}
\end{align}
Counting derivatives in the effective action, it is clear that both of the graviton vertex operators have to contribute two momenta each, while each of the gauge-field vertex operators has to contribute a single momentum since the amplitude contains the field strength rather than the gauge potential. This means that only specific pieces of the above vertex operators will contribute to the contractions.

First of all we see that we only need to consider contributions in the even spin-structure. The reason is that, upon writing $T^6=T^2\times T^2\times T^2$, there are six two-dimensional fermionic zero modes in the internal manifold (two for each torus) which, however, we cannot soak up all with the vertex operators we have at our disposal. Therefore, the odd-spin structure vanishes identically. 

For the sum over even spin-structures to be non-vanishing, all vertex operators have to contribute the fermion bilinear part in the left moving (supersymmetric) sector. This means that the graviton vertices have to provide an additional momentum coming from the exponential factor. This results in the following correlation function
\begin{align}
\couph{A}{B}=&\langle Z^1\bar{\partial} Z^2(x_1)\,\bar{Z}^2\bar{\partial} \bar{Z}^1(x_2)\rangle\cdot\langle\chi_1\chi_2(x_1)\,\bar{\chi}_1\bar{\chi}_2(x_2)\,\bar{\chi}_1\chi_2(x_3)\,\chi_1\bar{\chi}_2(x_4)\rangle\cdot\langle\bar{J}_A(x_3)\, \bar{J}_B(x_4)\rangle\,.\label{HetContractGauge}
\end{align}
As one can see, the correlator has split into three distinct contributions, which can be computed separately in a straight-forward manner
\begin{itemize}
\item Space-time fermion correlator:\\
Starting with the fermionic piece we have the following left-moving contribution
\begin{align}
\langle\chi_1&\chi_2(x_1)\,\bar{\chi}_1\bar{\chi}_2(x_2)\,\bar{\chi}_1\chi_2(x_3)\,\chi_1\bar{\chi}_2(x_4)\rangle=\nonumber\\
&=\sum_s\frac{\vartheta_s(x_1-x_2-x_3+x_4)\vartheta_s(x_1-x_2+x_3-x_4)\vartheta^2_s(0)\eta^{12}}{\vartheta^2(x_1-x_2)\vartheta^2(x_3-x_4)}=\eta^{12}\,,\label{ResFermCorr}
\end{align}
where $\vartheta$ are Jacobi theta-functions and $\eta$ is the Dedekind eta-function. In the last step, in order to perform the sum over all even spin structures $s$ we have used the Riemann summation identity. We thus find that the result is independent of the world-sheet positions $x_{i=1,2,3,4}$.
\item Space-time boson correlator:\\
As we have found no $x$-dependence in (\ref{ResFermCorr}), it follows that the full $x_1$ and $x_2$ dependence of $\couph{A}{B}$ is in the space-time bosonic correlator of (\ref{HetContractGauge}). Therefore, we can immediately move on to calculate the integrated expression
\begin{align}
\int d^2x_1\int d^2x_2\langle Z^1\bar{\partial} Z^2(x_1)\,\bar{Z}^2\bar{\partial} \bar{Z}^1(x_2)\rangle\,.\label{IntegBosCorr}
\end{align}
Fortunately, correlators of this type have already been studied before in \cite{Antoniadis:1995zn}. There the following generating functional was introduced and calculated explicitly
\begin{align}
G(\lambda,\tau,\bar{\tau})&=\sum_{g=1}^\infty\frac{1}{(g!)^2}\left(\frac{\lambda}{\tau_2}\right)^{2g}\langle\prod_{i=1}^g\int d^2x_i Z^1\bar{\partial}Z^2(x_i)\prod_{j=1}^{g}\int d^2y_j \bar{Z}^2\bar{\partial}\bar{Z}^1(y_j)\rangle=\nonumber\\
&=\left(\frac{2\pi i\lambda\bar{\eta}^3}{\bar{\vartheta}_1(\lambda,\bar{\tau})}\right)^2 e^{-\frac{\pi\lambda^2}{\tau_2}}\,.\label{GeneratingFunctional}
\end{align}
Thus, we can easily read off the answer for (\ref{IntegBosCorr}) by computing the coefficient of $\lambda^2$ in an expansion of $G(\lambda,\tau,\bar{\tau})$. To this end, following e.g. \cite{Marino:1998pg}, we can write
\begin{align}
\left(\frac{2\pi i\lambda\bar{\eta}^3}{\bar{\vartheta}_1(\lambda,\bar{\tau})}\right)^2e^{-\frac{\pi\lambda^2}{\tau_2}}=\sum_{k=0}^{\infty}\lambda^{2k}\mathcal{P}_{2k}(\hat{\bar{G}}_2,\ldots, \bar{G}^{2k})\,,\label{SpaceTimeCorrEisen}
\end{align}
where $\bar{G}_{2k}$ are particular normalisations of the Eisenstein series
\begin{align}
&\bar{G}_{2k}=2\zeta(2k) \bar{E}_{2k}\,,&&\text{and} &&\hat{\bar{G}}_2=2\zeta(2)\hat{\bar{E}}_2=2\zeta(2)\left(\bar{E}_2-\frac{3}{\pi\tau_2}\right)\,.
\end{align}
Since Eisenstein series will be very important for our further computations we have compiled some of their properties in appendix~\ref{App:Eisenstein}. Moreover, $\mathcal{P}_{2k}$ is a modular function (`almost' modular form) of weight $(0,2k)$
\begin{align}
\mathcal{P}_{2k}(\hat{\bar{G}}_2,\ldots, &\bar{G}_{2k})=-\mathcal{S}_k\left(\hat{\bar{G}}_2,\ldots,\frac{1}{k}\bar{G}_{2k}\right)\,,&&\text{with} &&\mathcal{S}_k(x_1,\ldots,x_k)=x_k+\ldots+\frac{x_1^k}{k!}\,,\nonumber
\end{align}
with $\mathcal{S}_k$ being the Schur polynomials. This particularly means
\begin{align}
&\eisen{2}=-\hat{\bar{G}}_2\,,&&\text{and} &&\eisen{4}=-\frac{1}{2}(\hat{\bar{G}}_2^2+\bar{G}_4)\,,
\end{align}
which entails for the correlator
\begin{align}
\int d^2x_1\int d^2x_2\langle Z^1\bar{\partial} Z^2(x_1)\,\bar{Z}^2\bar{\partial} \bar{Z}^1(x_2)\rangle=\eisen{2}(\hat{\bar{G}}_2)\,.
\end{align}
It is crucial to realise that although this correlator is a modular function of weight $(0,2)$ it is not an anti-holomorphic function due to the dependence of $\eisen{2}$ on $\hat{\bar{E}}_2$.
\item Current Correlator:\\
Finally, there is still the correlator of the right moving currents in (\ref{HetContractGauge}). Following \cite{Antoniadis:1992rq}, it is given by
\begin{align}
\langle\bar{J}_A(\bar{x}_3)\, \bar{J}_B(\bar{x}_4)\rangle&=P^R_AP^R_B-\frac{\delta_{AB}}{4\pi^2}\partial^2_{\bar{x}_3}\ln\bar{\vartheta}_1(\bar{x}_3-\bar{x}_4)\,,\label{RightMovingCorrelatorCurrents}
\end{align}
where $P^R_A$ is a right moving vector of the $\Gamma^{(6,22)}$-Narain lattice corresponding to the toroidal compactification. Since (\ref{RightMovingCorrelatorCurrents}) is the only dependence of $\mathcal{A}_{AB}^{\text{het}}$ on the insertion points $x_3$ and $x_4$, we can immediately consider the integrated version. To this end we make use of the fact that $\partial_{\bar{x}_3}\ln\bar{\vartheta}_1(\bar{x}_3-\bar{x}_4)+\frac{2\pi i}{\tau_2}\,\text{Im}(x_3-x_4)$ as a function of $x_3$ is periodic on the torus. Therefore, we can compute the integral
\begin{align}
\int d^2x_3&\int d^2x_4\langle\bar{J}_A(\bar{x}_3)\, \bar{J}_B(\bar{x}_4)\rangle=\tau_2^2\left[P^R_AP^R_B-\frac{\delta_{AB}}{4\pi\tau_2}\right]\,,\label{CurrentCorrelator}
\end{align}
where we have used the appropriate normalisation. In the final correlator this expression will be an insertion into the Siegel-Narain Theta-function of weight $(3,11)$, as we will see below.
\end{itemize}
\subsubsection{Modular Integral}\label{Sect:ModularIntegral}
Since from the above analysis we only found one non-vanishing contraction, we can easily reassemble the full amplitude. For this, we have to include the partition function of the space-time bosons and fermions, yielding a factor of $\eta^{-8}$, as well as the contribution of the internal CFT. The latter is a Siegel-Narain Theta-function with the insertions (\ref{CurrentCorrelator}). The full expression is then of the form
\begin{align}
\couph{A}{B}&\simeq\int \frac{d^2\tau}{\tau_2^3\bar{\eta}^{24}}\,\tau_2^4\,\eisen{2}\Gsum\left[P^R_AP^R_B-\frac{\delta_{AB}}{4\pi\tau_2}\right]\qq\,.\label{modularintegral}
\end{align}
As a simple check, we show in appendix~\ref{App:ModInvariance} that the integrand of this expression is indeed modular invariant. 

In order to compute this integral, we recall the following property of the function $\eisen{2}$
\begin{align}
\partial_\tau \eisen{4}=-\frac{i\pi}{2\tau_2^2}\,\eisen{2}\,.
\end{align}
Performing then an integration by parts we find (including the boundary contribution)
{\allowdisplaybreaks
\begin{align}
\couph{A}{B}=&-\frac{3}{\pi}\int \frac{d^2\tau}{\bar{\eta}^{24}}\,\tau_2^2\,\eisen{4}\Gsum\left[P^R_AP^R_B-\frac{\delta_{AB}}{4\pi\tau_2}\right]\qq+\nonumber\\
&+2\int \frac{d^2\tau}{\bar{\eta}^{24}}\,\tau_2^3\,\eisen{4}\Gsum\left[P^R_AP^R_B-\frac{\delta_{AB}}{4\pi\tau_2}\right](P^L)^2\qq-\nonumber\\
&-\frac{\delta_{AB}}{4\pi^2}\int\frac{d^2\tau}{\bar{\eta}^{24}}\tau_2\eisen{4}\Gsum\qq+\nonumber\\
&+\frac{1}{\pi}\int_{\partial\mathcal{F}}\frac{d\tau_1}{\bar{\eta}^{24}}\tau_2^3\,\eisen{4}\Gsum\left[P^R_AP_B^R-\frac{\delta_{AB}}{4\pi\tau_2}\right]\qq\,.
\end{align}}
Introducing covariant derivatives with respect to the moduli $D_{ij,A}$, which act in the following manner on the lattice momenta (for more details see \cite{Antoniadis:2006mr,Antoniadis:2007cw})
\begin{align}
&D_{ij,A}P^L_{kl}=\epsilon_{ijkl}P^R_A\,,&&\text{and} &&D_{ij,A}P^R_B=\frac{1}{2}\delta_{AB}P^L_{ij}\,,
\end{align}
we can rewrite this expression as:
{\allowdisplaybreaks
\begin{align}
\couph{A}{B}=\left(\frac{\epsilon^{ijkl}}{16\pi^2}\,D_{ij,A}D_{kl,B}+\frac{\delta_{AB}}{2\pi^2}\right)\mathcal{I}+\mathcal{I}_{AB}^{\text{bdy}}\,,
\end{align}}
where we have introduced the following shorthand notation for the modular integrals
\begin{align}
&\mathcal{I}=\int \frac{d^2\tau}{\bar{\eta}^{24}}\,\tau_2\,\eisen{4}\Gsum\qq\,,\label{F3ModIntegral}\\
&\mathcal{I}_{AB}^{\text{bdy}}=\frac{1}{\pi}\int_{\partial\mathcal{F}}\frac{d\tau_1}{\bar{\eta}^{24}}\tau_2^3\,\eisen{4}\Gsum\left[P^R_AP_B^R-\frac{\delta_{AB}}{4\pi\tau_2}\right]\qq\,.\label{F3ModIntegralBdy}
\end{align}
As they are written, (\ref{F3ModIntegral}) and (\ref{F3ModIntegralBdy}) are valid for a generic toroidal compactification of the heterotic string and as such depend on the full Narain-moduli space of the $T^6$-compactification. Besides being rather tedious to compute, these integrals are also not quite what we aim to do in this work. For latter applications it will be more convenient to go to a particular region in the moduli space where we can obtain certain simplifications. To be precise our choice is the following
\begin{itemize}
\item Upon writing the internal $T^6=T^2\times T^4$ we will consider the limit of large $T^4$ volume~$V$.
\item From all the moduli of the Narain lattice, we will consider the simplest case, namely that all 16 right moving Wilson lines are vanishing.
\end{itemize}
In this case, the lattice factorises in the following manner
\begin{align}
\Gamma^{(6,22)}\to \Gamma^{(2,2)}\oplus \Gamma^{(4,4)}\oplus \Gamma^{(0,16)}\,,\label{LatticeSplit}
\end{align}
with the large volume limit $\Gamma^{(4,4)}\sim \frac{V}{\tau_2^2}$. The third factor in (\ref{LatticeSplit}) will then just contribute the lattice sum, which is a modular form of weight $(0,8)$ and just depends on the gauge group of the heterotic string. At the one-loop level for $E_8\times E_8$ and $SO(32)$ it is explicitly given by (see e.g. \cite{Kiritsis:1997hj}) $\Gamma^{(0,16)}\simeq(\bar{E}_4)^2$. This moreover means that the only moduli dependence of $\couph{A}{B}$ stems from the $(T,U)$ moduli of the remaining $T^2$, which enters via the $\Gamma^{(2,2)}$ factor in (\ref{LatticeSplit}). Putting all contributions together, we obtain the following simplified expression for $\mathcal{I}$ and $\mathcal{I}_{AB}^{\text{bdy}}$
\begin{align}
&\mathcal{I}^{\text{sim}}=\int \frac{d^2\tau}{\tau_2}\,\frac{\eisen{4}(\bar{E}_4)^2}{\bar{\eta}^{24}}\GsumT\qq\,,\label{modularintegralTRUNC}\\
&\mathcal{I}^{\text{bdy-sim}}_{AB}=\frac{1}{\pi}\int_{\partial\mathcal{F}} d \tau_1\tau_2\,\frac{\eisen{4}(\bar{E}_4)^2}{\bar{\eta}^{24}}\GsumT\left[P^R_AP_B^R-\frac{\delta_{AB}}{4\pi\tau_2}\right]\qq\,.\label{modularbdyTRUNC}
\end{align}
We will first compute the boundary term $\mathcal{I}^{\text{bdy-sim}}_{AB}$ in (\ref{modularbdyTRUNC}). To this end we realize that the only contribution comes from the limit of $\tau_2\to \infty$. In this limit, however, the integral (\ref{modularbdyTRUNC}) is regularised by the presence of $\qq$, except for the point where $P^L=P^R=0$. Therefore we obtain 
\begin{align}
\mathcal{I}^{\text{bdy-sim}}_{AB}&=-\frac{\delta_{AB}}{4\pi^2}\lim_{\tau_2\to\infty}\int_{-1/2}^{1/2}d\tau_1\frac{\eisen{4}(\bar{E}_4)^2}{\bar{\eta}^{24}}=\frac{42\pi^2}{5}\,\delta_{AB}\,.\label{BdyTerm}
\end{align}
Finally we are left to calculate the integral $\mathcal{I}^{\text{sim}}$. As we can see, the advantage of all previous rewriting is that $\mathcal{I}^{\text{sim}}$ is now of the form 
\begin{align}
\int_{\mathcal{F}}\frac{d^2\tau}{\tau_2}\,\hat{F}(\bar{\tau})\,\Theta(\tau,\bar{\tau})\,,&&\text{with} &&\hat{F}=\frac{\eisen{4}(\bar{E}_4)^2}{\bar{\eta}^{24}}=\sum_{m\geq-1}\sum_{t=0}^2c(m,t)\bar{q}^m\tau_2^{-t}\,.\label{BorcherdsIntegral}
\end{align}
Here $\Theta$ is a Siegel-Narain theta-function and $\hat{F}$ is an `almost' anti-holomorphic modular function for which we have computed the first few $c(m,t)$ explicitly in appendix \ref{App:Eisenstein}. Integrals of the type (\ref{BorcherdsIntegral}) have been studied in \cite{Borcherds} (see also \cite{Marino:1998pg}) by developing further ideas of \cite{Harvey:1995fq} (for older works see also \cite{Dixon:1990pc}). Also in the present case the computation is along the lines of \cite{Marino:1998pg} and is performed in appendix~\ref{App:ExplicitTorusIntegral}. The result is in fact chamber-dependent, i.e. it depends on where exactly in the $(T,U)$-moduli space we are working. We have chosen to consider the region in which $T_2U_2$ becomes large, in which case we can finally give the full result\footnote{Notice that due to our simplifications the derivatives will be all anti-symmetrised combinations of $(T_1,T_2,U_1,U_2)$.}
\begin{align}
\couph{A}{B}=\left(\frac{\epsilon^{ijkl}}{16\pi^2}\,D_{ij,A}D_{kl,B}+\frac{3}{4\pi^2}\,\delta_{AB}\right)\mathcal{I}^{\text{sim}}+\mathcal{I}_{AB}^{\text{bdy-sim}}\,,
\end{align}
where we have found in (\ref{BdyTerm}), (\ref{GeneralBorcherdsInt})
\begin{align}
&\mathcal{I}_{AB}^{\text{bdy-sim}}=\frac{42\pi^3}{5}\,\delta_{AB}\,,&&\text{and} &&\mathcal{I}^{\text{sim}}=\frac{\mathcal{I}^{\text{sim}}_{K}}{\sqrt{2z_+^2}}+\mathcal{I}^{\text{sim}}_{\lambda=0}+\mathcal{I}^{\text{sim}}_{\lambda\neq 0}\,,
\end{align}
with the explicit expressions (\ref{ThetaTransform}), (\ref{IntDegRes}) and (\ref{IntNonDegRes}) for the chamber $T_2<U_2$
\begin{align}
&\frac{\mathcal{I}^{\text{sim}}_{K}}{\sqrt{2z_+^2}}=-\frac{16\pi^5}{3}\,U_2+2T_2\sum_{t=0}^2c(0,t)\frac{t!\zeta(2t+2)}{\pi^{t+1}}\,\left(\frac{T_2}{U_2}\right)^t\,,\\
&\mathcal{I}^{\text{sim}}_{\lambda=0}=c(0,0)\left[\gamma_E-\log\left(\pi T_2U_2\right)-2\log 2\right]+c(0,1)\frac{\zeta(3)}{\pi T_2U_2}+c(0,2)\frac{3\zeta(5)}{2\pi^2 T_2^2U_2^2}\,,\\
&\mathcal{I}^{\text{sim}}_{\lambda\neq 0}=\sum_{\lambda\neq 0} \sum_{t=0}^2\sum_{s=0}^tc(\lambda^2/2,t)(T_2U_2)^{-t}\frac{\left(\text{Im}(\alpha)\right)^{t-s}}{(4\pi)^s}\,\frac{(s+t)!}{s!(t-s)!}\,\text{Li}_{1+s+t}\left(e^{2\pi i\alpha}\right)\,.
\end{align}
This essentially concludes our calculation of $\mathcal{A}_{AB}^{\text{het}}$.
\subsection{One-Loop Graviphoton Amplitude in Heterotic String Theory}\label{Sect:OneLoopHetGraviphoton}
In addition to the gauge-field contribution, we can also consider whether there is a non-trivial coupling in which the gauge-fields are replaced by graviphotons. In fact, this is a non-trivial question for the following reason: As already explained in Section~\ref{Sect:Superfields} we are essentially considering 22+6 vector multiplets, the last six of which act as compensating multiplets. The gauge fields of the latter can -- via their equations of motion -- be expressed in terms of the 22 physical gauge fields as well as the graviphotons. In this way, all couplings which we can write down for the gauge fields might as well have partners containing graviphotons. 

To investigate this point, we can examine whether a four-point one-loop amplitude including the following vertex operators gives any non-vanishing contribution\\ 
\begin{center}
\begin{tabular}{|c|c|c|c|}\hline
\textbf{field} & \textbf{helicity} & \textbf{vertex} & \textbf{WS position}\\\hline
graviton & $R_{1212}$ & \parbox{6.9cm}{\vspace{0.1cm}$V^{(R)}(p_1)=\left(\partial Z^2-ip_1\chi^1\chi^2\right)\bar{\partial} Z^2e^{ip_1Z^1}$\vspace{0.1cm}} & $x_1$\\ \hline
graviton & $R_{\bar{1}\bar{2}\bar{1}\bar{2}}$ & \parbox{6.9cm}{\vspace{0.1cm}$V^{(R)}(\bar{p}_2)=\left(\partial \bar{Z}^1-i\bar{p}_2\bar{\chi}^2\bar{\chi}^1\right)\bar{\partial} \bar{Z}^1e^{i\bar{p}_2\bar{Z}^2}$\vspace{0.1cm}} & $x_2$\\ \hline
graviph. & $T_{\bar{1}2}$ & \parbox{6.6cm}{\vspace{0.1cm}$V^{(T)}(\bar{p}_1)=\left(\partial X-i\bar{p}_1\bar{\chi}^1\Psi\right)\bar{\partial}Z^2 e^{i\bar{p}_1\bar{Z}^1}$\vspace{0.1cm}} & $x_3$\\\hline
graviph. & $T_{1\bar{2}}$ & \parbox{6.6cm}{\vspace{0.1cm}$V^{(T)}(\bar{p}_2)=\left(\partial X-i\bar{p}_2\bar{\chi}^2\Psi\right)\bar{\partial} Z^1e^{i\bar{p}_2\bar{Z}^2}$\vspace{0.1cm}} & $x_4$\\\hline
\end{tabular}
\end{center}
${}$\\[10pt]
Here $X$ denotes the complex coordinate of the internal $T^2$ with $\Psi$ its supersymmetric partner. However, this amplitude is zero; to proof its vanishing it suffices to consider the fermion contribution. By inspection it is clear that the only possibility for contractions includes the fermionic correlator
\begin{align}
\langle\chi_1\chi_2(x_1)\bar{\chi}_1\bar{\chi}_2(x_2)\rangle=\sum_s\frac{\vartheta_s^2(x_1-x_2)\vartheta_s^2(0)}{\vartheta^2(x_1-x_2)}=0\,.
\end{align}
This establishes that there is no similar coupling involving graviphotons at one-loop. This result ties in with the expression for the higher-derivative couplings which we have obtained from harmonic superspace. Recalling the explicit component form (\ref{A2compExp}) we can see that the coupling only involves gauge fields from vector multiplets, but no graviphotons. Notice, however, that this analysis does not exclude such couplings appearing at higher loops (or non-perturbatively) in string theory. However, for this to happen, the corresponding harmonic superspace interaction will have to contain the dilaton in a non-trivial manner as we will discuss now. 
\subsection{Duality}
Before applying the results we have obtained so far to the study of entropy corrections in $\cN=4$ black holes, we would like to pause for a moment and discuss some aspects of duality covariance of the newly found higher derivative term (\ref{A2compExp}). The fact that this interaction only involves the $SO(22)$ gauge-fields $F_{A}^{\mu\nu}$ might lead to the suspicion that it breaks $SO(6,22)$ covariance. However, one way to see that this is not the case is to reformulate (\ref{A2compExp}) in the supergravity basis instead of the superstring basis (recall the discussion of section~\ref{Sect:Superfields}). In this basis, at the component level, we will find  
\begin{align}
S_{g=0}^{2,\text{SUGRA}}&\simeq\int d^4x R_{(+),\mu\nu\rho\tau}R_{(+)}^{\mu\nu\rho\tau} F_{(-),I,\sigma\lambda} F_{(-),J}^{\sigma\lambda} \int du\,{\mathcal{A}^{IJ}}_{\big|\theta=0}\,,\label{DualSUGRAframe}
\end{align}
with $I,J$ indices of $SO(6,22)$ and $\mathcal{A}^{IJ}$ an expression similar to (\ref{ShorthandCoupling}), which is a tensor-valued modular function of $SO(6,22)$. The expression (\ref{DualSUGRAframe}) is therefore manifestly $SO(6,22)$ covariant.

Switching to the superstring basis (which we have been using so far and which we will also use in the later sections) entails to replace the $SO(6)$ gauge fields $F^{\mu\nu}_{I=1,\ldots,6}$ by the graviphotons $T^{\mu\nu}_{ij}$. As for example explained in \cite{Antoniadis:1993ze}, this change of basis will involve the tree-level gauge-kinetic terms of the superstring action and therefore will also involve the heterotic dilaton. Thus, while the contribution of the $SO(22)$ gauge fields becomes precisely the term (\ref{A2compExp}), the corresponding contributions of the graviphotons will receive an extra dilaton dependence. These couplings will therefore not appear at the one-loop level in the superstring frame, but will only receive higher-loop or non-perturbative contributions. Notice that this is in perfect agreement with our explicit computation in section~\ref{Sect:OneLoopHetGraviphoton}. Only if these additional contributions are included, $SO(6,22)$ covariance will be restored in the superstring frame. 
\section{Entropy Corrections for Large Black Holes}\label{Sect:EntropyBlackLarge}
\setcounter{equation}{0}
After having studied the higher derivative couplings (\ref{A2compExp}) both from a superspace point of view and calculated them explicitly as heterotic string amplitudes, we now study whether they have any effect on the physics of (large) $\cN=4$ supersymmetric black holes.
\subsection{Spectrum and Charge Setup}\label{Sect:ChargeSetup}
So far we have been discussing an $\cN=4$ theory of $22$ physical vector multiplets coupled to the $\cN=4$ SUGRA multiplet. For computing the entropy of black holes, it will, however, be more useful to describe the theory in an $\cN=2$ language. In this case the $\cN=4$ SUGRA multiplet decomposes in the following manner
\begin{align}
[(2),4(3/2),6(1),4(1/2),(0)]&\longrightarrow[(2),2(3/2),(1)]\oplus2[(3/2),(1),(1/2)]\oplus[(1),2(1/2),(0)]\,.\label{SUGRAdecompose}
\end{align}
The right hand side corresponds to the $\cN=2$ SUGRA multiplet, two spin--$3/2$ multiplets and an $\cN=2$ vector multiplet. We recall that the scalar in this decomposition (i.e. the graviscalar in $\cN=4$) is identified with the heterotic dilaton in string theory. Each of the $\cN=4$ vector multiplets on the other hand side is decomposed as follows
\begin{align}
[(1),4(1/2),(0)]\longrightarrow[(1),2(1/2),(0)]\oplus[2(1/2),(0)]\,,\label{VectorDecompose}
\end{align}
where the right hand side corresponds to an $\cN=2$ vector and a hypermultiplet. 

The first step to describe a particular black hole in supergravity is to choose a particular setup of charges which it will carry. This means that we have to choose the black hole to be charged under some of the gauge fields inside the $\cN=2$ multiplets on the right hand side of (\ref{SUGRAdecompose}) and (\ref{VectorDecompose}) while the remaining multiplets will be truncated. Starting with the fields coming from the $\cN=4$ SUGRA multiplet in (\ref{SUGRAdecompose}), we choose the black hole to carry electric charges $q_1$ and $q_3$ with respect to the $\cN=2$ SUGRA (graviphoton) and the vector multiplet respectively and completely truncate the spin--$3/2$ multiplets. For the $\cN=4$ vector multiplets, we first recall that in the computation of the heterotic one-loop amplitude in Section~\ref{Sect:ModularIntegral} we have considered the limit of large $T^4$ volume. In this limit 20 of the $\cN=4$ physical vector multiplets get truncated and we are only left with those containing the $T$ and $U$ modulus of the remaining $T^2$ of the internal theory. From these -- under the decomposition (\ref{VectorDecompose}) -- we will keep the $\cN=2$ vector multiplets by choosing the black hole to carry magnetic charges $p_2$ and $p_4$ under the corresponding gauge fields while we will completely truncate  the hypermultiplets.

This choice of charges together with the large volume limit of $T^4$ makes it possible for us to make contact with the work of e.g. \cite{Sen:2005iz}, where black holes in heterotic string theory compactified on $\mathcal{M}\times S^1_{(1)}\times S^1_{(2)}$, with large volume of $\mathcal{M}$ (which is either $K3$ or $T^4$ or some orbifold thereof) were considered. As explained in \cite{Sen:2005iz}, in string theory the electric charges of the graviphotons can be interpreted as winding and momentum along the direction $S^1_{(1)}$ while the magnetic charges of the gauge fields correspond to Kaluza-Klein and H-monopole charge associated with $S^1_{(2)}$. In fact, to obtain the real physical quantum numbers $(n,w,N,W)$ (which are also quantised) the following redefinition is necessary
\begin{align}
&q_1=\frac{n}{2}\,,&&q_3=\frac{w}{2}\,,&&p_2=4\pi N\,,&&p_4=4\pi W\,.\label{ChargeRescaling}
\end{align}
In most of our calculations we will stick to the set $(q_1,q_3,p_2,p_4)$. Moreover, to match the assumptions we have made during the explicit computation of the one-loop amplitude and to guarantee a weakly coupled theory, we will have to impose the following hierarchy of charges
\begin{align}
q_1\gg q_3\gg p_2\gg p_4\ggg1\,.\label{chargeHier}
\end{align}
For completeness, let us also mention that the dual setup in type~II string theory compactified on $K3\times T^2$ corresponds to a D0-D4-D4-D4 brane configuration (see e.g. \cite{Dabholkar:2005dt}). There, the electric charges stem from D0-branes as well as a stack of D4-branes wrapping $K3$, while the magnetic charges correspond to the remaining two stacks of D4-branes which wrap $T^2\times \gamma_{1,2}$, where $\gamma_{1,2}$ are two 2-cycles inside $K3$.
\subsection{Entropy Function}\label{Sect:SecondDerEntrop}
We will now compute the entropy function \cite{Sen:2005wa,Sen:2005iz} for the black hole setup outlined in the previous subsection. We will work iteratively order by order in a derivative expansion of the effective action, starting with the tree-level one and assume large charges throughout.
\subsubsection{Ansatz for the Fields}\label{Sect:AnsatzFields}
Before considering the action, we have to make an ansatz for all fields of the theory in the vicinity of the horizon of the black hole. Starting with the metric we assume (following \cite{Sen:2005wa,Sen:2005iz}) that the near-horizon geometry is of the form $AdS_2\times S^2$ for which we make the ansatz
\begin{align}
ds^2=G_{\mu\nu}dx^\mu dx^\nu=v_1\left(-r^2dt^2+\frac{dr^2}{r^2}\right)+v_2(d\theta^2+\sin^2\theta d\varphi^2)\,.\label{metric}
\end{align}
Here $v_1$ and $v_2$ are two constants parameterising the radii of $AdS_2$ and $S^2$ respectively. We will determine both of them in the following. Concerning the scalar fields, after the truncation outlined in Section~\ref{Sect:ChargeSetup} we still have to deal with three of them: the heterotic dilaton (inside the $\cN=2$ SUGRA multiplet) and the $(T,U)$-moduli of $T^2$ (inside the two vector multiplets). We will make the following ansatz for them
\begin{align}
&e^{-2\Phi}=s\,,&&R_1=T_2U_2=r_1\,,&&R_2=\frac{T_2}{U_2}=r_2\,,
\end{align}
with $s$, $r_1$ and $r_2$ constants which need to be determined explicitly. Here we have chosen to follow \cite{Sen:2005iz} and consider the limit in which $T^2$ factorises into $S^1_{(1)}\times S^1_{(2)}$ with radii  $R_1$ and $R_2$ respectively. 

Finally for the gauge field strength tensors, following our outline of the charge setup in Section~\ref{Sect:ChargeSetup} we make the following ansatz
{\allowdisplaybreaks
\begin{align}
&F_{\mu\nu}^{(1)}=
 \left(
\begin{array}{cccc}
 0 & e_1 & 0 & 0 \\
 -e_1 & 0 & 0 & 0 \\
 0 & 0 & 0 & 0 \\
 0 & 0 & 0 & 0
\end{array}
\right)\,,&& &&F_{\mu\nu}^{(2)}= \left(
\begin{array}{cccc}
 0 & 0 & 0 & 0 \\
 0 & 0 & 0 & 0 \\
 0 & 0 & 0 & \frac{p_2 \sin\theta}{4 \pi } \\
 0 & 0 & -\frac{p_2 \sin\theta}{4 \pi } & 0
\end{array}
\right),\label{field1}\\
&\nonumber\\ 
&F_{\mu\nu}^{(3)}= \left(
\begin{array}{cccc}
 0 & e_3 & 0 & 0 \\
 -e_3 & 0 & 0 & 0 \\
 0 & 0 & 0 & 0 \\
 0 & 0 & 0 & 0
\end{array}
\right)\,, && &&F_{\mu\nu}^{(4)}= \left(
\begin{array}{cccc}
 0 & 0 & 0 & 0 \\
 0 & 0 & 0 & 0 \\
 0 & 0 & 0 & \frac{p_4 \sin\theta}{4 \pi } \\
 0 & 0 & -\frac{p_4 \sin\theta}{4 \pi } & 0
\end{array}
\right)\,,\label{field4}
\end{align}}
where $(p_2,p_4)$ are the magnetic charges respectively and $(e_1,e_3)$ are essentially the Legendre transforms of the electric charges $(q_1,q_3)$.
\subsubsection{Two Derivative Entropy Function}
We start by determining the entropy and near horizon geometry for a large black hole characterised by the charges $(q_1,q_3,p_2,p_4)$ in the classical limit. To this end, we consider the classical tree-level action given by (see \cite{Sen:2005iz})
\begin{align}
S^{\text{tree}}=\frac{1}{32\pi}\int d^4x&\sqrt{-G} e^{-2\phi} \bigg[R+4\partial_\mu\phi\partial^\mu\phi -r_1^{-2}\partial_\mu r_1\partial^\mu r_1-r_2^{-2}\partial_\mu r_2\partial^\mu r_2-\nonumber\\
-&r_1^2F^{(1)}_{\mu\nu}F^{(1),\mu\nu}-r_2^2F^{(2)}_{\mu\nu}F^{(2),\mu\nu}-r_1^{-2}F^{(3)}_{\mu\nu}F^{(3),\mu\nu}-r_2^{-2}F^{(4)}_{\mu\nu}F^{(4),\mu\nu}\bigg],\label{TreeLevelAction}
\end{align}
where $R$ is the Ricci scalar computed from the space-time metric $G_{\mu\nu}$ with determinant $G$. This action gives rise to the following entropy function
\begin{align}
\mathcal{E}_{(2)}=2\pi (e_1q_1+e_3q_3)-\frac{\pi  s (v_1-v_2)}{2}-\frac{\pi sv_2}{2v_1}\left( e_1^2r_1^2+\frac{e_3^2}{r_1^2}\right)+\frac{sv_1}{32\pi v_2}\left(p_2^2r_2^2+\frac{p_4^2}{r_2^2}\right)\,,\label{TreeLevelEntropyFunction}
\end{align}
whose extremum with respect to the parameters $(v_1,v_2,s,r_1,r_2)$ is the leading order entropy.\footnote{For a pedagogical outline of the entropy-function formalism see e.g. \cite{Sen:2007qy}. Notice moreover that in some cases in the literature (e.g.\cite{Sen:2007qy,Sen:2005iz}) it has been shown to be useful to perform a suitable $SO(6,22)$-rotation of the charges such that a number of gauge fields will decouple at the attractor point. We have chosen not to perform such a rotation in the following but we will directly extremize the entropy function thereby directly obtaining the attractor values of all fields.} A quick computation reveals that the extremum is situated at 
\begin{align}
&v_1=v_2=\frac{p_2p_4}{4\pi^2}\,,&& &&s=\frac{8\pi\sqrt{q_1q_3}}{\sqrt{p_2p_4}}\,,&& &&r_1=\sqrt{\frac{q_1}{q_3}}\,,\label{nearhorsol1}\\ 
&r_2=\sqrt{\frac{p_4}{p_2}}\,,&& &&e_1=\frac{q_3\sqrt{p_2p_4}}{4\pi\sqrt{q_1q_3}}\,,&& &&e_3=\frac{q_1\sqrt{p_2p_4}}{4\pi\sqrt{q_1q_3}}\,,\label{nearhorsol2}
\end{align}
from which the entropy follows to be
\begin{align}
\mathcal{S}_{(2)}=\sqrt{q_1q_3p_2p_4}=2\pi\sqrt{nwNW}\,.\label{BHentropyLARGE}
\end{align}
This result has already been obtained in \cite{Sen:2005iz}. We will now consider corrections to this result due to the 4th-order higher derivative terms, similar to \cite{Sen:2005iz}.
\subsubsection{Four Derivative Entropy Function}\label{Sect:FourthDerEntrop}
The first correction to the entropy will stem from four derivative terms in the effective action. The full tree-level contribution to these terms in heterotic string theory is given by the dimensional reduction of a manifestly covariant term in six dimensions \cite{Metsaev:1987zx,Hull:1987pc} together with the gravitational Chern-Simons term. However, it was proven in \cite{Sahoo:2006pm} that the contribution of these terms to the black hole entropy is the same with the one obtained from the four-dimensional Gauss-Bonnet term. Since the computations are much simpler in this case, we will simply follow \cite{Sen:2005iz} and add the Gauss-Bonnet term to the tree-level action (\ref{TreeLevelAction}) (for related computations in a non-supersymmetric setup see \cite{Olea:2005gb}):
\begin{align}
\Delta S_{GB}=-\frac{3}{16\pi^2}\int d^4x \ln\left(2s |\eta(a+is)|^4\right) \left(R_{\mu\nu\rho\tau}R^{\mu\nu\rho\tau}-4R_{\mu\nu}R^{\mu\nu}+R^2\right)\,,\label{GaussBonnet}
\end{align}
where $a$ is the axion field. Using the same ansatz as in Section \ref{Sect:SecondDerEntrop} we can write the modified entropy function in a straight-forward manner
\begin{align}
\mathcal{E}_{(4)}=&2 \pi (e_1 q_1+e_3 q_3)-\frac{\pi s (v_1-v_2)}{2}-\frac{\pi sv_2}{2v_1}\left( e_1^2r_1^2+\frac{e_3^2}{r_1^2}\right)+\frac{sv_1}{32\pi v_2}\left(p_2^2r_2^2+\frac{p_4^2}{r_2^2}\right)-\nonumber\\
&-\frac{3}{8\pi}  \ln\left(2s|\eta(a+is)|^4\right)\,.
\end{align}
The extremum with respect to the axion is fixed by
\begin{align}
\frac{\partial \mathcal{E}_{(4)}}{\partial a}=-\frac{3}{4\pi}\left(\frac{\eta'(a+is)}{\eta(a+is)}-\frac{\eta'(-a+is)}{\eta(-a+is)}\right)=0\,,
\end{align}
which has a solution at $a=0$. Introducing the shorthand notation
\begin{align}
\zeta(s):=-\frac{3}{32\pi^2}\ln\left(2s|\eta(is)|^4\right)\,,\label{ZetaDef}
\end{align}
the entropy function takes the form
\begin{align}
\mathcal{E}_{(4)}=&2 \pi (e_1 q_1+e_3 q_3)-\frac{\pi s (v_1-v_2)}{2}-\frac{\pi sv_2}{2v_1}\left( e_1^2r_1^2+\frac{e_3^2}{r_1^2}\right)+\frac{sv_1}{32\pi v_2}\left(p_2^2r_2^2+\frac{p_4^2}{r_2^2}\right)+4\pi\zeta(s)\,.\label{RedEntropR2}
\end{align}
Extremising this function with respect to $(v_1,v_2,r_1,r_2)$ is straight-forward, as it only requires solving polynomial equations. The answer (in terms of the remaining variable $s$) is given by
\begin{align}
&v_1=v_2=\frac{p_2p_4}{8\pi^2}+\frac{8q_1q_3}{s^2}\,,&&\text{and} &&r_1=\sqrt{\frac{q_1}{q_3}}\,,&&\text{and}&&r_2=\sqrt{\frac{p_4}{p_2}}\,.\label{nearhorsolR2pre}
\end{align}
As we can see, using relation (\ref{chargeHier}) it follows that $T_2U_2\gg 1$ and $T_2\ll U_2$ which matches our assumptions of appendix~\ref{App:ExplicitTorusIntegral}. The solution for the Legendre transformed electric charges is given by
\begin{align}
&e_1=\frac{2q_3}{s}\,,&&\text{and} &&e_3=\frac{2q_1}{s}\,.
\end{align}
However, extremising the entropy function (\ref{RedEntropR2}) also with respect to the dilaton $s$ is more involved, due to the presence of the non-trivial function $\zeta(s)$ (see (\ref{ZetaDef})). We therefore need to find a way of approximating the equation. To this end, we make the following ansatz for $s$ based on (\ref{nearhorsol1})
\begin{align}
s=\frac{8\pi\sqrt{q_1q_3}}{\sqrt{p_2p_4}}+x_s\,,
\end{align}
with $x_s$ a function of the charges of order $\mathcal{O}(q^{-2},p^{-2})$. With this ansatz, we have to solve
\begin{align}
\frac{\sqrt{p_2^3p_4^3}}{64\pi^2\sqrt{q_1q_3}}\,x_s+4\pi\zeta'\left(\frac{8\pi\sqrt{q_1q_3}}{\sqrt{p_2p_4}}\right)+\mathcal{O}(q^{-2},p^{-2})=0\,,
\end{align}
which has the solution
\begin{align}
x_s=-\frac{256\pi^3\sqrt{q_1q_3}}{\sqrt{p_2^3p_4^3}}\,\zeta'\left(\frac{8\pi\sqrt{q_1q_3}}{\sqrt{p_2p_4}}\right)+\mathcal{O}(q^{-4},p^{-4})\,.
\end{align}
Therefore, the final result to leading order in the charges is given by
\begin{align}
&v_1=v_2=\frac{p_2p_4}{4\pi^2}+8\zeta'\left(\frac{8\pi\sqrt{q_1q_3}}{\sqrt{p_2p_4}}\right)+\mathcal{O}(q^{-2},p^{-2})\label{SolR21}\,,\\
&s=\frac{8\pi\sqrt{q_1q_3}}{\sqrt{p_2p_4}}-\frac{256\pi^3\sqrt{q_1q_3}}{\sqrt{p_2^3p_4^3}}\,\zeta'\left(\frac{8\pi\sqrt{q_1q_3}}{\sqrt{p_2p_4}}\right)+\mathcal{O}(q^{-4},p^{-4})\,,\\
&e_1=\frac{q_3\sqrt{p_2p_4}}{4\pi\sqrt{q_1q_3}}+\frac{64q_3\pi^2}{p_2p_4}\,\zeta'\left(\frac{8\pi\sqrt{q_1q_3}}{\sqrt{p_2p_4}}\right)+\mathcal{O}(q^{-3},p^{-3})\,,\\
&e_3=\frac{q_1\sqrt{p_2p_4}}{4\pi\sqrt{q_1q_3}}+\frac{64q_1\pi^2}{p_2p_4}\,\zeta'\left(\frac{8\pi\sqrt{q_1q_3}}{\sqrt{p_2p_4}}\right)+\mathcal{O}(q^{-3},p^{-3})\,,
\end{align}
while the moduli $r_1$ and $r_2$ remain the same as in (\ref{nearhorsol1}) and (\ref{nearhorsol2})
\begin{align}
&r_1=\sqrt{\frac{q_1}{q_3}}+\mathcal{O}(q^{-4},p^{-4})\,,&&\text{and} &&r_2=\sqrt{\frac{p_4}{p_2}}+\mathcal{O}(q^{-4},p^{-4})\,.\label{SolR25}
\end{align}
Inserting this result into (\ref{RedEntropR2}), we get the following expression for the black hole entropy
\begin{align}
\mathcal{S}_{(4)}&=\sqrt{q_1q_3p_2p_4}+4\pi\zeta\left(\frac{8\pi\sqrt{q_1q_3}}{\sqrt{p_2p_4}}\right)+\mathcal{O}(q^{-2},p^{-2})=\nonumber\\
&=2\pi\sqrt{nwNW}+4\pi\zeta\left(\sqrt{\frac{nw}{NW}}\right)+\mathcal{O}(n^{-2},w^{-2},N^{-2},W^{-2})\,.\label{BHentropyR2}
\end{align}
With this result we are now ready to include the effect of the six-derivative terms.
\subsubsection{Six Derivative Entropy Function}
In order to reduce writing to a minimum, we use the following shorthand notation for the effective coupling $\couph{A}{B}$ of (\ref{ShorthandCoupling})
\begin{align}
\couph{A}{B}=\xi_{AB}(r_1,r_2)\,.
\end{align}
The precise moduli dependence has been computed in Section~\ref{Sect:OneLoopHeterotic}. Notice, since $\couph{A}{B}$ is a one-loop amplitude, $\xi$ is independent of the dilaton $s$. With this and using the same ansatz for the fields in the near horizon area of the black hole as in Section~\ref{Sect:AnsatzFields}, the contribution of the six-derivative term (\ref{A2compExp}) reads
\begin{align}
\mathcal{E}_{(6)}=&\,2 \pi (e_1 q_1+e_3 q_3)-\frac{\pi s (v_1-v_2)}{2}-\frac{\pi sv_2}{2v_1}\left( e_1^2r_1^2+\frac{e_3^2}{r_1^2}\right)+\frac{sv_1}{32\pi v_2}\left(p_2^2r_2^2+\frac{p_4^2}{r_2^2}\right)+\nonumber\\
&+4\pi  \zeta(s)-\frac{4 (v_1^2+v_2^2)}{v_1 v_2^3}\,(\xi_{AB} (r_1,r_2)p^Ap^B)\,.\label{Entrop62}
\end{align}
Here we have combined the two magnetic charges into a vector of the form
\begin{align}
p^A=\left(\begin{array}{c}p_2 \\ p_4\end{array}\right)\,.\label{VectCharge}
\end{align}
Extremisation of (\ref{Entrop62}) with respect to $(e_1,e_3)$ can be performed analytically yielding
\begin{align}
&e_1=\frac{2q_1v_1}{sr_1^2v_2}\,,&&\text{and}&&e_3=\frac{2q_3r_1^2v_1}{s2v_2}\,.
\end{align}
For the remaining parameters $(v_1,v_2,s,r_1,r_2)$ an analytic solution for the full entropy function turns out to be quite difficult to obtain, mostly due to the complicated functions $\zeta(s)$ and $\xi(r_1,r_2)$. We therefore again proceed by searching for an approximated solution. To this end, we make the following ansatz based on (\ref{SolR21})--(\ref{SolR25})
\begin{align}
&v_1=\frac{p_2p_4}{4\pi^2}+8\zeta'\left(\frac{8\pi\sqrt{q_1q_3}}{\sqrt{p_2p_4}}\right)+x_1\,,&&v_2=\frac{p_2p_4}{4\pi^2}+8\zeta'\left(\frac{8\pi\sqrt{q_1q_3}}{\sqrt{p_2p_4}}\right)+x_2\,,\label{HigherOrderAnsatzv1}
\end{align}
\begin{align}
s=\frac{8\pi\sqrt{q_1q_3}}{\sqrt{p_2p_4}}-\frac{256\pi^3\sqrt{q_1q_3}}{\sqrt{p_2^3p_4^3}}\,\zeta'\left(\frac{8\pi\sqrt{q_1q_3}}{\sqrt{p_2p_4}}\right)+x_s\,,\label{HigherOrderAnsatzs}
\end{align}
\begin{align}
&r_1=\sqrt{\frac{q_1}{q_3}}+x_{r_1}\,,&&r_2=\sqrt{\frac{p_4}{p_2}}+x_{r_2}\,,\label{ansatzSOL2}
\end{align}
where $(x_{1},x_{2})$ are assumed to be of order $\mathcal{O}(q^{-2},p^{-2})$ and $(x_s,x_{r_1},x_{r_2})$ of order $\mathcal{O}(q^{-4},p^{-4})$ in the charges.

With this ansatz, we can extremise the entropy function (\ref{Entrop62}) to leading order, finally obtaining the following result 
{\allowdisplaybreaks
\begin{align}
v_1=&\frac{p_2p_4}{4\pi^2}+8\zeta'\left(\frac{8\pi\sqrt{q_1q_3}}{\sqrt{p_2p_4}}\right)-\frac{2048\pi^3\sqrt{q_1q_3}}{\sqrt{p_2^3p_4^3}}\,\zeta'\left(\frac{8\pi\sqrt{q_1q_3}}{\sqrt{p_2p_4}}\right)\,\zeta''\!\left(\frac{8\pi\sqrt{q_1q_3}}{\sqrt{p_2p_4}}\right)\,,\label{SOLR2full1}\\
&\nonumber\\
v_2=&\frac{p_2p_4}{4\pi^2}+8\zeta'\left(\frac{8\pi\sqrt{q_1q_3}}{\sqrt{p_2p_4}}\right)-\frac{2048\pi^3\sqrt{q_1q_3}}{\sqrt{p_2^3p_4^3}}\,\zeta'\left(\frac{8\pi\sqrt{q_1q_3}}{\sqrt{p_2p_4}}\right)\,\zeta''\!\left(\frac{8\pi\sqrt{q_1q_3}}{\sqrt{p_2p_4}}\right)-\nonumber\\*
&-\frac{64\pi^2p^Ap^B}{\sqrt{q_1q_3}\sqrt{p_2^3p_4^3}}\,\xi_{AB}\!\left(\sqrt{\frac{q_1}{q_3}},\sqrt{\frac{p_4}{p_2}}\right)\,,\\
&\nonumber\\
s=&\frac{8\pi\sqrt{q_1q_3}}{\sqrt{p_2p_4}}-\frac{256\pi^3\sqrt{q_1q_3}}{\sqrt{p_2^3p_4^3}}\,\zeta'\left(\frac{8\pi\sqrt{q_1q_3}}{\sqrt{p_2p_4}}\right)+\frac{2048\pi^5p^Ap^B}{p_2^3p_4^3}\,\xi_{AB}\!\left(\sqrt{\frac{q_1}{q_3}},\sqrt{\frac{p_4}{p_2}}\right)+\nonumber\\*
&+\frac{12288\pi^5\sqrt{q_1q_3}}{\sqrt{p_2^5p_4^5}}\,\left(\zeta'\left(\frac{8\pi\sqrt{q_1q_3}}{\sqrt{p_2p_4}}\right)\right)^2+\frac{65536\pi^6q_1q_3}{p_2^3p_4^3}\,\zeta'\left(\frac{8\pi\sqrt{q_1q_3}}{\sqrt{p_2p_4}}\right)\,\zeta''\!\left(\frac{8\pi\sqrt{q_1q_3}}{\sqrt{p_2p_4}}\right)+\nonumber\\
&+\mathcal{O}(q^{-6},p^{-6})\,,\\
&\nonumber\\
r_1=&\sqrt{\frac{q_1}{q_3}}+\mathcal{O}(q^{-6},p^{-6})\,,\hspace{2cm}\text{and} \hspace{2cm}
r_2=\sqrt{\frac{p_4}{p_2}}+\mathcal{O}(q^{-6},p^{-6})\,.\label{SOLR2full5}
\end{align}}
Inserting this into (\ref{Entrop62}), we find for the entropy
\begin{align}
\mathcal{S}_{(6)}=&\sqrt{q_1q_3p_2p_4}+4\pi\zeta\left(\frac{8\pi\sqrt{q_1q_2}}{\sqrt{p_2p_4}}\right)-\frac{512\pi^4\sqrt{q_1q_2}}{\sqrt{p_2^3p_4^3}}\left(\zeta'\left(\frac{8\pi\sqrt{q_1q_3}}{\sqrt{p_2p_4}}\right)\right)^2-\nonumber\\
&-\frac{128\pi^4p^Ap^B}{p_2^2p_4^2}\,\xi_{AB}\!\left(\sqrt{\frac{q_1}{q_3}},\sqrt{\frac{p_4}{p_2}}\right)+\mathcal{O}(q^{-4},p^{-4})\,.\label{BHentropyR2full}
\end{align}
With the physical quantum numbers (\ref{ChargeRescaling}) this becomes
\begin{align}
\mathcal{S}_{(6)}=&2\pi\sqrt{nwNW}+4\pi\zeta\left(\sqrt{\frac{nw}{NW}}\right)-\frac{4\pi\sqrt{nw}}{\sqrt{N^3W^3}}\left(\zeta'\left(\sqrt{\frac{nw}{NW}}\right)\right)^2-\nonumber\\
&-\frac{8\pi^2N^AN^B}{N^2W^2}\,\xi_{AB}\!\left(\sqrt{\frac{n}{w}},\sqrt{\frac{W}{N}}\right)+\mathcal{O}(n^{-4},w^{-4},N^{-4},W^{-4})\,.
\end{align}
Here we have also combined $(N,W)$ into $N^A$ in a similar fashion as in (\ref{VectCharge}). Notice that this correction is precisely of the expected order in the charges. Moreover, we see that there are in fact two correction terms. The first one, which depends on $\zeta$, is just the higher order correction from the Gauss-Bonnet term (\ref{GaussBonnet}). The last term, on the other hand, is proportional to $\xi$ and therefore is a contribution stemming from the six-derivative term (\ref{A2compExp}).
\section{Entropy Corrections for Small Black Holes}\label{Sect:EntropyBlackSmall}
\setcounter{equation}{0}
As we have seen in the previous section, in the case of large black holes, i.e. those which already classically have a non-vanishing horizon, the topological terms (\ref{A2compExp}) give only a subleading contribution to the entropy. One can now ask what the situation is in the case of small black holes for which a non-vanishing horizon is only provided by higher derivative terms in the effective supergravity action. In particular, it would be interesting to understand whether there are black holes for which the first non-trivial contribution to the entropy is provided by (\ref{A2compExp}). In this Section we would like to take a first step into this direction by considering two special cases. 
\subsection{Charge Setup}
We wish to consider particular limits of the charge setup discussed in Section~\ref{Sect:ChargeSetup}, namely we want to calculate the entropy in the case that we set to zero two out of the four charges $(q_1,q_3,p_2,p_4)$. Obviously we cannot simply apply this limit to the final result (\ref{BHentropyR2full}) since we have assumed throughout the computation in Section~\ref{Sect:SecondDerEntrop} that all charges are very large and we therefore have to perform the computations from scratch. To be more precise, with respect to the four charges $(q_1,q_3,p_2,p_4)$ there are two possible limits which we are interested in, namely vanishing magnetic charges $p_2=p_4=0$ and vanishing electric charges $q_1=q_3=0$. 

The first option has already been studied in \cite{Dabholkar:2004yr} on the type~II side. As higher derivative correction terms the 
topological $R^2$ interaction for a $K3\times T^2$ compactification was added.\footnote{This is the first term of the series of the topological $R^2T^{2g-2}$ couplings \cite{Antoniadis:1993ze}. However, in backgrounds with $\cN=4$ supersymmetry, only the term for $g=1$ yields a non-zero contribution.} It was proven explicitly that this term is not only responsible for the black hole to obtain a finite-size horizon but that the entropy calculated for this setup matches the result of the microstate counting to all orders in the large charge expansion. Put it differently, the $R^2$-interaction already captures the complete entropy of the black hole. It is therefore an interesting check for the consistency of our computations to see that the six derivative topological term (\ref{A2compExp}) does not modify this result. That this is indeed the case is quick to see. According to our discussion in Section~\ref{Sect:ChargeSetup} the two remaining gauge fields in this setting correspond to two graviphotons. However, in this case, as explicitly calculated in Section~\ref{Sect:OneLoopHetGraviphoton}, there is no contribution of the type (\ref{A2compExp}) and the result of \cite{Dabholkar:2004yr} is not modified.

One is therefore left to consider the second option, namely setting $q_1=q_3=0$. For this case we will now compute the entropy function including the fourth derivative Gauss-Bonnet term (\ref{GaussBonnet}) as well as the sixth-derivative coupling (\ref{A2compExp}).
\subsection{Entropy Function}
\subsubsection{Four Derivative Entropy Function}
We will use the same ansatz for the fields in the near-horizon region of the black hole as in Section~\ref{Sect:AnsatzFields}, however, with $q_1$ and $q_3$ set to zero. In this case extremisation of the tree-level entropy function yields a vanishing entropy. We therefore immediately proceed to include the Gauss-Bonnet term (\ref{GaussBonnet}). To be explicit, we will use (\ref{ProdRepDedekind}) to expand the function $\zeta(s)$ introduced in (\ref{ZetaDef}) in powers of $s$ in the following manner
\begin{align}
\zeta(s):=-\frac{3}{32\pi^2}\ln\left(2s|\eta(is)|^4\right)=\frac{s}{32\pi}-\frac{3}{32\pi^2}\,\log(2s)+\ldots\,.\label{ZetaExpand}
\end{align}
Here the dots stand for exponentially suppressed terms, whose contributions we are not interested in. With this explicit expression, the fourth-derivative entropy function takes the form
\begin{align}
\mathcal{E}_{(4)}^{\text{small}}=-\frac{\pi s(v_1-v_2)}{2}+\frac{sv_1}{32\pi v_2}\left(p_2^2r_2^2+\frac{p_4^2}{r_2^2}\right)+\frac{s}{8}-\frac{3}{8\pi}\,\log(2s)\,.\label{EntrFct4DerSmall}
\end{align}
The extremum of this function is at the point
\begin{align}
&v_1=v_2=\frac{p_2p_4}{8\pi^2}\,,&&\text{and}&&r_2=\sqrt{\frac{p_4}{p_2}}\,,&&\text{and} &&s=\frac{6}{p_2p_4+2\pi}=\frac{6}{p_2p_4}-\frac{12\pi}{p_2^2p_4^2}+\mathcal{O}(p^{-6})\,.\label{Solution4DerSmall}
\end{align}
Note that since $\mathcal{E}_{(4)}^{\text{small}}$ is independent of the modulus $r_1$ its extremisation does not provide a value for it. Therefore, to this order in the charges, the entropy function formalism does not provide an attractor equation for $r_1$. Nevertheless, inserting (\ref{Solution4DerSmall}) into (\ref{EntrFct4DerSmall}), we obtain the entropy of the small black hole
\begin{align}
\mathcal{S}_{(4)}^{\text{small}}&=\frac{3}{8\pi}\log\left(\frac{p_2p_4+2\pi}{12}\right)+\frac{3}{8\pi}=\frac{3}{8\pi}\log\left(\frac{p_2p_4}{12}\right)+\frac{3}{8\pi}+\mathcal{O}(p^{-2})=\nonumber\\
&=\frac{3}{8\pi}\log\left(\frac{4\pi^2 NW}{3}\right)+\frac{3}{8\pi}+\mathcal{O}(N^{-2},W^{-2})\,.
\end{align}
As we can see, the entropy depends logarithmically on the charges. The reason for this is that the first non-trivial contribution essentially comes from the second (logarithmic) term in (\ref{ZetaExpand}), while the first term taken alone would still give a vanishing entropy.
\subsubsection{Six Derivative Entropy Function}
We now want to include also the sixth derivative topological terms (\ref{A2compExp}) for a twofold reason. On the one hand, we want to see whether it also contributes to the entropy of this black hole (although maybe in a subdominant way) and on the other hand, we want to check whether it allows to fix the value of the remaining modulus $r_1$. The modified entropy function is given by the expression
\begin{align}
\mathcal{E}_{(6)}^{\text{small}}=&-\frac{\pi s(v_1-v_2)}{2}+\frac{sv_1}{32\pi v_2}\left(p_2^2r_2^2+\frac{p_4^2}{r_2^2}\right)+\frac{s}{8}-\frac{3}{8\pi}\,\log(2s)-\nonumber\\
&-\frac{4(v_1^2+v_2^2)p^Ap^B}{v_1v_2^3}\,\xi_{AB}(r_1,r_2)\,.\label{EntrFct6DerSmall}
\end{align}
Extremizing this expression is rather difficult due to the presence of the complicated function $\xi_{AB}(r_1,r_2)$. We will therefore apply the same strategy as in Section~\ref{Sect:SecondDerEntrop} and linearise the equations around the solution (\ref{Solution4DerSmall}) by making the ansatz
\begin{align}
&v_1=\frac{p_2p_4}{8\pi^2}+x_1\,,&&v_2=\frac{p_2p_4}{8\pi^2}+x_2\,,&&s=\frac{6}{p_2p_4}-\frac{12\pi}{p_2^2p_4^2}+x_s\,,&&r_2=\sqrt{\frac{p_4}{p_2}}+x_{r_2}\,.
\end{align}
Here we assume the following scaling behaviour of the corrections
\begin{align}
&x_1\sim x_2=\mathcal{O}(p^0)\,,  &&x_s=\mathcal{O}(p^{-4})\,,&&r_1=\mathcal{O}(p^0)\,,&&x_{r_2}=\mathcal{O}(p^{-2})\,.
\end{align}
Extremizing (\ref{EntrFct6DerSmall}) to leading order in the charges amounts for $r_1$ to solve
\begin{align}
\frac{\partial}{\partial r_1}\left[p^Ap^B\xi_{AB}\left(r_1,\sqrt{\frac{p_4}{p_2}}\right)\right]=0\,,
\end{align}
whose solution $r_1^{(0)}$ therefore corresponds to the attractor value. Extremizing then $\mathcal{E}_{(6)}^{\text{small}}$ for the remaining quantities $(x_1,x_2,x_s,x_{r_2})$ yields the following next-to-leading order solution
\begin{align}
&v_1=\frac{p_2p_4}{8\pi^2}+\frac{1024\pi^2p^Ap^B}{3p_2p_4}\,\xi_{AB}\left(r_1^{(0)},\sqrt{\frac{p_4}{p_2}}\right)+\mathcal{O}(p^{-2})\,,\\
&v_2=\frac{p_2p_4}{8\pi^2}+\mathcal{O}(p^{-2})\,,\\
&s=\frac{6}{p_2p_4}-\frac{12\pi}{p_2^2p_4^2}+\mathcal{O}(p^{-6})\,,\\
&r_2=\sqrt{\frac{p_4}{p_2}}+\frac{1024\pi^5p^Ap^B}{3p_2^3p_4}\,\xi_{AB}^{(0,1)}\left(r_1^{(0)},\sqrt{\frac{p_4}{p_2}}\right)+\mathcal{O}(p^{-4})\,.
\end{align}
Here $\xi_{AB}^{(0,1)}$ denotes the first derivative of $\xi_{AB}$ with respect to the second argument. Reinserting this solution into (\ref{EntrFct6DerSmall}) we obtain the corrected entropy
\begin{align}
\mathcal{S}_{(6)}^{\text{small}}&=\frac{3}{8\pi}\log\left(\frac{p_2p_4}{12}\right)+\frac{3}{8\pi}+\frac{3}{4p_2p_4}-\frac{2048\pi^4p^Ap^B}{4p_2^2p_4^2}\,\xi_{AB}\left(r_1^{(0)},\sqrt{\frac{p_4}{p_2}}\right)+\mathcal{O}(p^{-4})=\nonumber\\
&=\frac{3}{8\pi}\log\left(\frac{4\pi^2 NW}{3}\right)+\frac{3}{8\pi}+\frac{3}{64\pi^2 NW}-\frac{32\pi^2N^AN^B}{N^2W^2}\,\xi_{AB}\left(r_{1}^{(0)},\sqrt{\frac{W}{N}}\right)+\nonumber\\
&\hspace{1cm}+\mathcal{O}(N^{-4},W^{-4})\,.
\end{align}
Since this result depends on $\xi_{AB}$, it follows that the sixth-derivative terms (\ref{A2compExp}) indeed yield a non-trivial contribution to the entropy. However, looking more precisely, this contribution is in fact subdominant with respect to the contribution coming from the Gauss-Bonnet term (\ref{GaussBonnet}).
\section{Conclusions}\label{Sect:Conclusions}
\setcounter{equation}{0}
In this work we have studied the effects of a particular topological six-derivative term on the entropy of black holes. We have explicitly calculated this term as a one-loop contribution in the effective heterotic string action, performing also the integral over the modular parameter of the world-sheet torus. 

In the case of large black holes, this term yields a non-vanishing correction to the entropy of the order $\mathcal{O}(p^{-2},q^{-2})$. For small black holes, we have studied two different setups: Black holes carrying only charges with respect to two graviphotons do not receive any corrections at all. This is in perfect agreement with the literature (see e.g. \cite{Dabholkar:2004yr}) where it has been shown that the entropy of such black holes is already captured by the topological fourth-derivative $R^2$ effective action coupling. 

On the other hand, for small black holes which are only charged with respect to two physical gauge fields, the leading contribution to the entropy also comes from $R^2$ terms (e.g. the Gauss-Bonnet combination), however not from the tree-level expression but rather from higher logarithmic corrections. In this setup the topological sixth-derivative corrections are still suppressed being of order $\mathcal{O}(p^{-2},q^{-2})$. However, they are responsible for lifting certain flat directions in the moduli space of the entropy function, thereby providing attractor values for some of the scalar fields involved.

It would be very interesting to compare our macroscopic results with some results obtained from state-counting. This would allow us to obtain a microscopic interpretation of the entropy in the setup we considered. Microscopic computations up to order $-2$ in the charges have recently been performed in \cite{Banerjee:2008ky}. There, it was speculated about the nature of higher derivative terms in the effective action which would be responsible for these entropy corrections on the macroscopic side. In this spirit, the term we have discussed in this paper seems to be a good candidate for this task. However, as far as we can see, in order to be able to make a precise comparison between our macroscopic calculations and the microscopic results of \cite{Banerjee:2008ky} it seems necessary to taken into account non-local terms in the effective action which arise upon integrating out massless degrees of freedom. We leave this study for further work.
\section*{Acknowledgements}
It is a pleasure to thank Atish Dabholkar, Sergio Ferrara, Finn Larsen, Boris Pioline, Frank Saueressig, Ashoke Sen and Emery Sokatchev for enlightening discussions. S.H. would like to thank the Laboratoire de Physique Th\'eorique et Hautes Energies (LPTHE) in Paris for kind hospitality during the initial stage of this work. The work of I.A. was supported in part by the European Commission under the ERC Advanced Grant 226371 and the contract PITN-GA-2009-237920 and in part by the CNRS grant GRC APIC PICS 3747. The research of S.H. was supported by the Swiss National Science Foundation.
\appendix
\section{Modular Functions and Eisenstein Series}\label{App:Eisenstein}
\renewcommand{\theequation}{\Alph{section}.\arabic{equation}}
\setcounter{equation}{0}
Since they play a major role throughout the heterotic one-loop computation in Section~\ref{Sect:OneLoopHeterotic}, we will compile some useful identities and formulas for Eisenstein series in this appendix. 

The functions $G_{2k}$ appearing in the generating functional (\ref{SpaceTimeCorrEisen}) are the canonically defined Eisenstein series
\begin{align}
G_{2k}(\tau)=\sum_{\text{\tiny $\begin{array}{c}m,n=-\infty \\ mn\neq 0\end{array}$}}^{\infty}(m\tau+n)^{-2k}\,.
\end{align}
In this work we will also use a different normalisation of the Eisenstein series
\begin{align}
E_{2k}(q)=\frac{G_{2k}(\tau)}{2\zeta(2k)}=1+c_{2k}\sum_{n=1}^\infty\sigma_{2k-1}(n)q^n\,,
\end{align}
where $q=e^{2i\pi\tau}$, $\sigma_{k}(n)$ is the divisor function (i.e. the sum of the $k$-th powers of the integer divisors of $n$), and
\begin{align}
c_{2k}=\frac{(2\pi i)^{2k}}{(2k-1)!\zeta(2k)}\,.
\end{align}
For latter use we give the explicit $q$-expansion of the first few $E_{2k}$
\begin{align}
&E_{2}(q)=1-24q-72q^2-96q^3-168q^4-144q^5+\ldots\,,\\
&E_{4}(q)=1+240q+2160q^2+6720q^3+17520q^4+30240q^5+\ldots\,,\\
&E_{6}(q)=1-504q-16632q^2-122976q^3-532728q^4-1575504q^5+\ldots\,,\\
&E_{8}(q)=1+480q+61920q^2+1050240q^3+7926240q^4+37500480q^5+\ldots\,.
\end{align}
For $k>1$, $G_{2k}$ (and $E_{2k}$) are modular functions of weight $2k$. However, $G_2$  picks up an additional shift term under modular transformations, instead of which we introduce 
\begin{align}
\hat{G}_2=2\zeta(2)\hat{E}_2=2\zeta(2)\left(E_2-\frac{3}{\pi\tau_2}\right)\,.
\end{align}
The additional term cancels precisely the shift rendering $\hat{G}_2$ a modular function of weight two, however, at the expense of being no longer purely holomorphic.

Using moreover the expansion of the Dedekind function
\begin{align}
\eta(\tau)&=q^{\frac{1}{24}}\left[1+\sum_{n=1}^{\infty}(-1)^n\left(q^{n(3n-1)/2}+q^{n(3n+1)/2}\right)\right]=q^{\frac{1}{24}}\left(1-q-q^2+q^5+q^7+\ldots\right)=\nonumber\\
&=q^{\frac{1}{24}}\prod_k^\infty\left(1-q^k\right)\,,\label{ProdRepDedekind}
\end{align}
we are finally in a position to determine the first few expansion coefficients $c(m,t)$ in (\ref{BorcherdsIntegral}). They are given by
\begin{align}
&c(-1,0)=-\frac{\pi^4}{15}\,,&&c(-1,1)=\frac{\pi^3}{3}\,,&&c(-1,2)=-\frac{\pi^2}{2}\,,\\
&c(0,0)=-\frac{168\pi^4}{5}\,,&&c(0,1)=160\pi^3\,,&&c(0,2)=-252\pi^2\,,\\
&c(1,0)=-\frac{24828\pi^4}{5}\,,&&c(1,1)=20532\pi^3\,,&&c(1,2)=-36882\pi^2\,,\\
&c(2,0)=-\frac{612352\pi^4}{3}\,,&&c(2,1)=-\frac{888320\pi^3}{3}\,,&&c(2,2)=-1347520\pi^2\,,\\
&c(3,0)=-7320798\pi^4\,,&&c(2,1)=-5094930\pi^3\,,&&c(2,2)=-27377865\pi^2\,,\\
&c(4,0)=-\frac{1002596352\pi^4}{5}\,,&&c(3,1)=-245568768\pi^3\,,&&c(3,2)=-389320128\pi^2\,.
\end{align}
\section{Modular Invariance}\label{App:ModInvariance}
\renewcommand{\theequation}{\Alph{section}.\arabic{equation}}
\setcounter{equation}{0}
In this appendix we check modular invariance of the integrand of (\ref{modularintegral}). To this end, we will separately check invariance under the two generators of the modular group $\tau\to\tau+1$ and $\tau\to-\frac{1}{\tau}$. Indeed, the first one can be checked in a straight-forward manner. Using the fact that $\hat{\bar{E}}_2$ and $\bar{\eta}^{24}$ are respectively invariant under the shift, we find that under $\tau\to\tau+1$
\begin{align}
\couph{A}{B}\to\int \frac{d^2\tau}{\bar{\eta}^{24}}\,\tau_2\,\hat{\bar{E}}_2\Gsum\left[P^R_AP^R_B-\frac{\delta_{AB}}{4\pi\tau_2}\right]\qq e^{\pi i\left[(P^L)^2-(P^R)^2\right]}\,.
\end{align}
However, since $\Gamma^{(6,22)}$ is a self-dual lattice, the additional phase in the lattice sum is in fact one. We are therefore left to consider the transformation $\tau\to-\frac{1}{\tau}$. Using that the Dedekind functions transforms as $\bar{\eta}^2\to\bar{\tau}\bar{\eta}^2$ we find for $\couph{A}{B}$
\begin{align}
\couph{A}{B}\to \tilde{\mathcal{A}}_{AB}^{\text{het}}=\int \frac{d^2\tau}{\bar{\eta}^{24}}\,\frac{\tau_2\hat{\bar{E}}_2}{\tau^3\bar{\tau}^{13}}\Gsum\left[P^R_AP^R_B-\frac{\delta_{AB}\tau\bar{\tau}}{4\pi\tau_2}\right]e^{-\frac{\pi i}{\tau}(P^L)^2}e^{\frac{\pi i}{\bar{\tau}}(P^R)^2}\,.
\end{align}
For this expression we now perform a Poisson resummation
\begin{align}
\tilde{\mathcal{A}}_{AB}^{\text{het}}=\int \frac{d^2\tau}{\bar{\eta}^{24}}&\,\frac{\tau_2\hat{\bar{E}}_2}{\tau^3\bar{\tau}^{13}}\sum_{(X^L,X^R)\in\Gamma^{(6,22)}}\int d^6P^L\int d^{22}P^R\left[P^R_AP^R_B-\frac{\delta_{AB}\tau\bar{\tau}}{4\pi\tau_2}\right]e^{-\frac{\pi i}{\tau}(P^L)^2}e^{\frac{\pi i}{\bar{\tau}}(P^R)^2}\cdot\nonumber\\
&\cdot e^{2\pi i(X^L\cdot P^L)}e^{-2\pi i(X^R\cdot P^R)}\,.\nonumber
\end{align}
Transforming to new coordinates $Y^L_{ij}=(P^L_{ij}-\tau X^L_{ij})$ and $Y^R_A=(P^R_A-\bar{\tau}X^R_A)$ and evaluating explicitly the Gaussian integrals we obtain the expressions:
\begin{align}
\tilde{\mathcal{A}}_{AB}^{\text{het}}=\int \frac{d^2\tau}{\bar{\eta}^{24}}&\,\frac{\tau_2\hat{\bar{E}}_2}{\tau^3\bar{\tau}^{13}}\sum_{(X^L,X^R)\in\Gamma^{(6,22)}}\left[\tau^3\bar{\tau}^{13}X^R_AX^R_B-\frac{\delta_{AB}\tau^3\bar{\tau}^{13}}{4\pi\tau_2}\right]q^{\frac{1}{2}(X^L)^2}\bar{q}^{\frac{1}{2}(X^R)^2}\,.
\end{align}
This amounts to $\tilde{\mathcal{A}}_{AB}^{\text{het}}=\couph{A}{B}$ which finishes the proof of modular invariance of (\ref{modularintegral}).
\section{Torus Integral via Lattice Reduction}\label{App:ExplicitTorusIntegral}
In this appendix we explicitly compute the modular integral $\mathcal{I}^{\text{sim}}$ of (\ref{modularintegralTRUNC}), where we will mainly follow \cite{Marino:1998pg,Borcherds}. The first step is to reduce the $\Gamma^{(2,2)}$ unimodular lattice to a $\Gamma^{(1,1)}$ sublattice. For this, we start by writing $\Gamma^{(2,2)}$ in the form\footnote{We will use bold-face letters to denote lattice vectors.}
\begin{align}
\Gamma^{(2,2)}=H(-1)\oplus H(1)=\langle {\bf e_1},{\bf f_1}\rangle_{\mathbb{Z}}\oplus\langle {\bf e_2},{\bf f_2}\rangle_{\mathbb{Z}}\,, 
\end{align}
with $({\bf e_1},{\bf f_1})=-({\bf e_2},{\bf f_2})=-1$ the only non-vanishing inner products. In addition to the lattice we also have an isometry $P:\,\Gamma^{(2,2)}\otimes \mathbb{R}\longrightarrow \mathbb{R}^{2,2}$, whose projection to $\mathbb{R}^{2,0}$ and $\mathbb{R}^{0,2}$ will be called $P_\pm$ respectively. Explicitly, for a given vector ${\mathbf{\lambda}}$, we have
\begin{align}
&P^L=P_-({\bf \lambda})=\frac{1}{\sqrt{2T_2U_2}} \left(n_1+n_2\bar{T}+m_2U+m_1\bar{T}U\right)\,,\\
&P^R=P_+({\bf \lambda})=\frac{1}{\sqrt{2T_2U_2}}\left(n_1+n_2T+m_2U+m_1TU\right)\,.
\end{align}
In order now to perform a lattice reduction, we pick a primitive null-vector ${\bf z}$ inside $\Gamma^{(2,2)}$ alongside with another vector ${\bf z'}$, such that $({\bf z},{\bf z'})=1$. A natural choice for this is to pick ${\bf z}={\bf e_1}$ and ${\bf z'}=-{\bf f_1}$. With this vector we can define a new lattice 
\begin{align}
K=\left(\Gamma^{(2,2)}\cap {\bf z^\perp}\right)/\mathbb{Z}{\bf z}\,,
\end{align}
which is of signature $(1,1)$. Here $\mathbb{Z}{\bf z}$ stands for all integer multiples of the null-vector ${\bf z}$. In $K$ we will define new projections $\tilde{P}_\pm$. To this end, we denote the projections of ${\bf z}$ in the old lattice as ${\bf z_\pm}=P_\pm({\bf z})$, for which we find explicitly
\begin{align}
z_+^2=|{\bf z_+}|^2=\frac{1}{2T_2U_2}\,.\label{FirstLattRedVector}
\end{align}
In fact, in order for the lattice reduction to be valid, this expression needs to be small (see \cite{Borcherds}), which entails that we need to restrict to a region in moduli space, where $T_2U_2\gg 1$.

With ${\bf z_\pm}$ we can decompose
\begin{align}
&\mathbb{R}^{2,0}=\langle {\bf z_+}\rangle\oplus\langle {\bf z_+}\rangle^\perp\,,&&\text{and} &&\mathbb{R}^{0,2}=\langle {\bf z_-}\rangle\oplus\langle {\bf z_-}\rangle^\perp\,.
\end{align}
The reduced projections $\tilde{P}_\pm$ will then be the projections onto the orthogonal complement $\langle {\bf z_+}\rangle^\perp$ and $\langle {\bf z_-}\rangle^\perp$, respectively. They are given in terms of the old projections $P_\pm$ in the following manner
\begin{align}
\tilde{P}_\pm({\bf \lambda})=P_\pm({\bf \lambda})-\frac{(P_\pm({\bf\lambda}),{\bf z_\pm})}{z_\pm^2}\,{\bf z_\pm}\,.
\end{align}
With this, the lattice momenta in the new lattice are given by
\begin{align}
&\tilde{P}^L=\frac{1}{\sqrt{2T_2U_2}}\left(n_2\bar{T}+m_2U\right)\,,&&\text{and} &&\tilde{P}^R=\frac{1}{\sqrt{2T_2U_2}}\left(n_2T+m_2U\right)\,.
\end{align}
Following \cite{Marino:1998pg}, it particularly follows for a vector $\lambda\in K$
\begin{align}
&\tilde{P}_+(\lambda)=\text{Im}(\tilde{P}^R)\,&&\text{and} &&\tilde{P}_-(\lambda)=\text{Im}(\tilde{P}^L)\,.\label{ProjRedLatt}
\end{align}
For latter use, let us also introduce the following vector in $K\otimes \mathbb{R}$
\begin{align}
\mu=-{\bf z'}+\frac{{\bf z_+}}{2z_+^2}+\frac{{\bf z_-}}{2z_-^2}\,.
\end{align}
At this point we can use the final result of \cite{Borcherds,Marino:1998pg}: The theta-transform $\mathcal{I}^{\text{sim}}$ is given as a sum of three terms 
\begin{align}
&\mathcal{I}^{\text{sim}}=\frac{\mathcal{I}^{\text{sim}}_{K}}{\sqrt{2z_+^2}}+\mathcal{I}^{\text{sim}}_{\lambda=0}+\mathcal{I}^{\text{sim}}_{\lambda\neq0}\,.\label{GeneralBorcherdsInt}
\end{align}
Here $\mathcal{I}^{\text{sim}}_{K}$ is another theta-transform, however, in the reduced lattice $K$. Moreover, the remaining two contributions are given by
\begin{align}
&\mathcal{I}_{\lambda=0}^{\text{sim}}=\sqrt{\frac{2}{z_+^2}}\sum_{n>0}\sum_tc(0,t)\left(\frac{\pi n^2}{2z_+^2}\right)^{-\epsilon-t-1/2}\Gamma\left(t+\frac{1}{2}+\epsilon\right)_{\big|\epsilon=0}\,, \label{BorchDeg}\\
&\mathcal{I}_{\lambda\neq0}^{\text{sim}}=\sqrt{\frac{2}{z_+^2}}{\sum_{\lambda\in K}}'\sum_{n>0}e^{(n\lambda,\mu)}\sum_t2c(\lambda^2/2,t)\left(\frac{n}{2|z_+||\tilde{P}_+(\lambda)|}\right)^{-t-1/2}\!\mathcal{K}_{-t-1/2}\left(\frac{2\pi n|\tilde{P}_+(\lambda)|}{|z_+|}\right) \label{BorchNonDeg}
\end{align}
where the prime on the sum over $\lambda$ in (\ref{BorchNonDeg}) means that the zero-vector is excluded and $\mathcal{K}_{-t-1/2}$ is a modified Bessel function of second kind. Equation (\ref{BorchDeg}) must be understood as the constant piece of an analytic Laurent expansion in $\epsilon$.

In the following we discuss all three contributions in detail.
\begin{itemize}
\item Reduced theta-transform (degenerate orbit)\\
In order to compute the left-over theta-transform, we perform another lattice reduction to arrive at the trivial lattice. For this, we pick the vectors ${\bf \tilde{z}}={\bf e_2}$ and ${\bf \tilde{z}'}={\bf f_2}$ which particularly yields
\begin{align}
\tilde{z}_+^2=|{\bf \tilde{z}_+}|^2=\frac{T_2}{2U_2}\,.
\end{align}
Notice that with this choice we are working in the patch $T_2<U_2$. Exchanging ${\bf \tilde{z}}$ and ${\bf \tilde{z}'}$ will bring us to the patch $T_2>U_2$. All results will be exactly the same upon the exchange $T_2\longleftrightarrow U_2$. With this reduction, we are left with two contributions
\begin{align}
\frac{\mathcal{I}^{\text{sim}}_{K}}{\sqrt{2z_+^2}}=&\frac{1}{2\sqrt{z_+^2\tilde{z}_+^2}}\int_{\mathcal{F}}\frac{d^2\tau}{\tau_2^2}\,\frac{\eisen{4}(\bar{E}_4)^2}{\bar{\eta}^{24}}+\nonumber\\
&+\frac{1}{\sqrt{z_+^2\tilde{z}_+^2}}\sum_{t=0}^2c(0,t)\frac{2^{t+1}\Gamma(t+1)}{\pi^{t+1}}\,(\tilde{z}_+^2)^{t+1}\zeta(2t+2)\,.\label{FullRedThetaTransform}
\end{align}
The first term is an integral over the fundamental domain, which can nevertheless be evaluated directly
\begin{align}
\int_{\mathcal{F}}\frac{d^2\tau}{\tau_2^2}\,\frac{\eisen{4}(\bar{E}_4)^2}{\bar{\eta}^{24}}&=-\frac{1}{2}\int_{\mathcal{F}}\frac{d^2\tau}{\tau_2^2}\left[\frac{\bar{\hat{G}}_2^2(\bar{E}_4^2)}{\bar{\eta}^{24}}+\frac{\bar{G}_4(\bar{E}_4^2)}{\bar{\eta}^{24}}\right]=-\frac{16\pi^5}{3}\,,
\end{align}
where we have used \cite{Lerche:1988np} (see also \cite{Marino:1998pg})
\begin{align}
\int_{\mathcal{F}}\frac{d^2\tau}{\tau_2^2}\left(\hat{G}_2(\tau)\right)^n F(\tau)=\frac{1}{\pi(n+1)}\left[(G_2(\tau))^{n+1}F(\tau)\right]_{q^0}\,.
\end{align}
Inserting this expression into (\ref{FullRedThetaTransform}) we obtain
\begin{align}
\frac{\mathcal{I}^{\text{sim}}_{K}}{\sqrt{2z_+^2}}=-\frac{16\pi^5}{3}\,U_2+2T_2\sum_{t=0}^2c(0,t)\frac{t!\zeta(2t+2)}{\pi^{t+1}}\,\left(\frac{T_2}{U_2}\right)^t\,.\label{ThetaTransform}
\end{align}

\item $\lambda=0$ contribution to the non-degenerate orbit\\
Next we will discuss the contribution to the non-degenerate orbit given in (\ref{BorchDeg}). First of all, following \cite{Marino:1998pg}, the sum over $n$ can be analytically continued into a Riemann zeta-function, leaving
\begin{align}
\mathcal{I}_{\lambda=0}^{\text{sim}}=\sqrt{\frac{2}{z_+^2}}\sum_tc(0,t)\left(\frac{\pi }{2z_+^2}\right)^{-\epsilon-t-1/2}\zeta(1+2t+2\epsilon)\Gamma\left(t+\frac{1}{2}+\epsilon\right)_{\big|\epsilon=0}\,.
\end{align}
Extraction of the constant piece in the $\epsilon$-expansion can be done in a straight-forward way yielding
\begin{align}
\mathcal{I}^{\text{sim}}_{\lambda=0}&=c(0,0)\left[\gamma_E-\log\left(\pi T_2U_2\right)-2\log 2\right]+c(0,1)\frac{\zeta(3)}{\pi T_2U_2}+c(0,2)\frac{3\zeta(5)}{2\pi^2 T_2^2U_2^2}\,,\label{IntDegRes}
\end{align}
where $\gamma_E$ is the Euler-Mascheroni constant. 
\item $\lambda\neq0$ contribution to the non-degenerate orbit\\
Finally, we are left to deal with the contribution (\ref{BorchNonDeg}) which we compute following a very similar computation in \cite{Marino:1998pg}. To this end, we choose a parameterisation of the vector $\lambda\in K$ of the form $\lambda=n_2 {\bf e_2}+m_2 {\bf f_2}$. In addition, we introduce the following shorthand notation
\begin{align}
\alpha=\frac{1}{2}\text{Re}(n_2T+m_2U)+i|\text{Im}(n_2T+m_2U)|\,,
\end{align}
upon which we find 
\begin{align}
|\tilde{P}_+(\lambda)|=\text{Im}(\tilde{P}^R)=\frac{1}{\sqrt{2T_2U_2}}|\text{Im}(n_2T+m_2U)|=\frac{\text{Im}(\alpha)}{\sqrt{2T_2U_2}}\,.
\end{align}
Using moreover the relation $({\bf z_+},\lambda)=\sqrt{z_+^2}\,\text{Re}(\tilde{P}^R)$ for $\lambda\in K$ we derive the following expression
\begin{align}
\mathcal{I}^{\text{sim}}_{\lambda\neq0}&=\sqrt{\frac{2}{z_+^2}}\sum_{\lambda\neq 0}\sum_{n=1}^\infty \sum_{t=0}^22c(\lambda^2/2,t)\left(\frac{nT_2U_2}{\text{Im}(\alpha)}\right)^{-t-\frac{1}{2}}\,e^{2\pi in\text{Re}(\alpha)}\, \mathcal{K}_{-t-\frac{1}{2}}(2\pi n\text{Im}(\alpha))\,.\nonumber
\end{align}
Using $\mathcal{K}_{-s}=\mathcal{K}_{s}$ together with its precise definition we can also write
\begin{align}
\mathcal{I}^{\text{sim}}_{\lambda\neq0}&=\sum_{\lambda\neq 0} \sum_{t=0}^2\sum_{s=0}^tc(\lambda^2/2,t)(T_2U_2)^{-t}\frac{\left(\text{Im}(\alpha)\right)^{t-s}}{(4\pi)^s}\,\frac{(s+t)!}{s!(t-s)!}\,\text{Li}_{1+s+t}\left(e^{2\pi i\alpha}\right)\,,\label{IntNonDegRes}
\end{align}
where $\text{Li}_{1+s+t}\left(e^{2\pi\alpha}\right)$ denotes the polylogarithm. 
\end{itemize}

\end{document}